\documentclass[prb,superscriptaddress,showpacs,floatfix,tightenlines,10pt,twocolumn]{revtex4}
\usepackage{amssymb}
\usepackage{amsmath}
\usepackage{graphicx}
\renewcommand{\Re}{\operatorname{Re}}
\renewcommand{\Im}{\operatorname{Im}}

\begin{document}

\title{Electromagnetic absorption and Kerr effect in quantum Hall
ferromagnetic states of bilayer graphene}
\author{R. C\^{o}t\'{e}}
\affiliation{D\'{e}partement de physique, Universit\'{e} de Sherbrooke, Sherbrooke, Qu%
\'{e}bec, J1K 2R1, Canada}
\author{Manuel Barrette}
\affiliation{D\'{e}partement de physique, Universit\'{e} de Sherbrooke, Sherbrooke, Qu%
\'{e}bec, J1K 2R1, Canada}
\author{\'{E}lie Bouffard}
\affiliation{D\'{e}partement de physique, Universit\'{e} de Sherbrooke, Sherbrooke, Qu%
\'{e}bec, J1K 2R1, Canada}

\begin{abstract}
In a quantizing magnetic field, the chiral two-dimensional electron gas in
Landau level $N=0$ of bilayer graphene goes through a series of phase
transitions at integer filling factors $\nu \in \left[ -3,3\right] $ when
the strength of an electric field applied perpendicularly to the layers is
increased. At filling factor $\nu =3,$ the electron gas can described by a
simple two-level system where layer and spin degrees of freedom are frozen.
The gas then behaves as an orbital quantum Hall ferromagnet. A
Coulomb-induced Dzyaloshinskii-Moriya term in the orbital pseudospin
Hamiltonian is responsible for a series of transitions first to a Wigner
crystal state and then to a spiral state as the electric field is increased.
Both states have a non trivial orbital pseudospin texture. In this work, we
study how the phase diagram at $\nu =3$ is modified by an electric field
applied in the plane of the layers and then derive several experimental
signatures of the uniform and nonuniform states in the phase diagram. In
addition to the transport gap, we study the electromagnetic absorption and
the Kerr rotation due to the excitations of the orbital pseudospin-wave
modes in the broken-symmetry states.
\end{abstract}

\date{\today }
\maketitle

\section{INTRODUCTION}

In a Bernal-stacked graphene bilayer (BLG), electrons behave as a chiral
two-dimensional Fermi gas.\cite{BilayerReview} When quantized by a strong
perpendicular magnetic field and when Coulomb interaction is considered, a
rich set of phase transitions occurs in Landau level $N=0$\cite%
{BarlasReview,ReneSU8,Gorbar} as well as in higher Landau levels $\left\vert
N\right\vert >0.$\cite{ReneIsing} The diversity of phases is greater in
level $N=0$ which has an extra orbital degree of freedom $n=0,1$ in addition
to the valley $\xi =\pm 1$ (for valleys $K_{\pm }$) and spin $\sigma =\pm 1$
quantum numbers shared by all the other Landau levels. At a fixed magnetic
field, and in a transverse electric field $E_{\bot },$ a different sequence
of phase transitions is obtained at each filling factor $\nu \in \left[ -3,3%
\right] $ in $N=0$ when a potential difference $\Delta _{B}=E_{\bot }d$ (or
bias) between the two layers with separation $d$ is increased.

The various phases driven by the bias $\Delta _{B}$ can be described as
quantum Hall ferromagnets (QHF's). In the pseudospin language, the two spin,
valley, and orbital indices are mapped into a $1/2$ spin, valley pseudospin
and orbital pseudospin respectively. In the simplest tight-binding model
where only in-plane and inter-plane hoppings $\gamma _{0}$ and $\gamma _{1}$
are considered (see below for a more precise definition of these terms) and
in the absence of Coulomb interaction, Zeeman and bias couplings, all eight
states in $N=0$ are degenerate. However, additional couplings such as the $%
\gamma _{4}$ hopping term breaks the orbital degeneracy while a Zeeman or a
bias term break the spin and the valley degeneracy respectively. The
degeneracy can also be spontaneously broken by the Coulomb interaction
leading to different types of quantum Hall ferromagnetic states. At the
Hartree-Fock level, it has been shown that Coulomb interaction completely
lifts the degeneracy of the $N=0$ octet leading to the formation of seven
new plateaus in the Hall conductivity$.$\cite{BarlasPRL,ReneSU8} These
plateaus have been detected experimentally.\cite{Experiences}

Most of the research done so far on the QHF's in $N=0$ has considered
uniform states. But, at special filling factors $\nu =1,3$, a sequence of
phase transitions involving uniform and nonuniform states is also possible.%
\cite{ReneSU8,ReneOrbital,ReneHelical} The nonuniform states occur in a
region of bias where the system can be described as an orbital QHF i.e.
where the electrons collectively condense into a linear combination of the $%
n=0$ and $n=1$ orbitals. The sequence of transitions for $\nu =3$ is
represented schematically in Fig. 1 below. It involves spiral and crystal
states where the orbital pseudospin is modulated in pseudospin space as well
as uniform states with and without orbital coherence. The rotation of the
pseudospin is induced by an effective Dzyaloshinskii-Moriya (DM) interaction
due to the Coulomb exchange term in the Hamiltonian. There is no
spin-orbital coupling in the 2DEG. The sequence of transitions in Fig. 1 is
similar to that uncovered in helical magnets such as MnSi and Fe$_{1-x}$Co$%
_{x}$Si\cite{HelicalMagnets}. In these systems, however, the transitions are
induced by changing the magnetic field.

In Refs. \onlinecite{ReneOrbital} and \onlinecite{ReneHelical}, some of us
have studied several aspects of the phase diagram in Fig. 1: the order
parameters, the density of states, the band structure, the collective
excitations. In the present work, we study in more detail possible transport
and optical experimental signatures of the different phases in this diagram
that have, so far, not been detected experimentally. Optical spectroscopy
has been used extensively to study graphene and multilayer graphene
structures (see Ref. \onlinecite{PotemskiReview} for an overview of the
subject). This includes absorption from Landau level transitions and Faraday
and Kerr effects (optical polarization rotation of the transmitted and
reflected wave respectively). The Faraday rotation, in particular, was shown
to be very large in graphene.\cite{GiantFaraday} Optical methods can also be
used to study broken-symmetry states.\cite{CalculKerr} Indeed, the
signatures of several gapped states on the optical conductivity of bilayer
graphene have been studied in some detail in Ref. \onlinecite{GorbarFaraday}
but this did not include the nonuniform states that we considered in this
article.

Optical transitions between non-interacting states in BLG\cite%
{PotemskiReview} must satisfy the selection rule $\left\vert N\right\vert
\rightarrow $ $\left\vert N\right\vert \pm 1$ and are naturally classified
as right and left circularly polarized transitions because of the change in
orbital momentum between the two levels involved in the transition. The same
division occurs in the simple case of $\nu =3$ in $N=0$ where the system can
effectively be mapped into a two-level system with orbital degrees of
freedom $n=0,1$ and where spin and valley indices are polarized. (The phase
diagram is slightly more complex for $\nu =1$ and includes states with spin
coherence.) In the left of the phase diagram in Fig. 1 (i.e. for bias $%
\Delta _{B}<\Delta _{M}$ where $\Delta _{M}$ defines the middle of the
spiral state), electrons occupy mostly the $n=0$ states and optical
absorption in all phases occurs for transition from $n=0$ to $n=1.$ Just the
opposite is true in the right of the phase diagram (for bias $\Delta
_{B}>\Delta _{M}$) where electrons occupy mostly the $n=1$ states. As we
will show, a spontaneous uniform orbital QHF does not lead to absorption at
finite frequency since its pseudospin-wave mode is gapless. The other
phases, however, have modes that can be excited with right or left
circularly polarized light. It should thus be possible to distinguish the
phases by the optical absorption due to their collective modes. In Fig. 1,
pairs of states (or "conjugate states") on each side of $\Delta _{M}$ share
the same dispersion of their collective modes as well as other physical
properties such as the transport gap. They absorb electromagnetic radiation
at the same frequency although with different intensities and from opposite
circular polarizations of an incident electromagnetic wave. The Kerr
rotation also show noticeably different behaviors for conjugate biases: the
polarization rotates in opposite directions and the amplitude of the
rotation is much larger for bias $\Delta _{B}>\Delta _{M}$. The symmetry of
the phase diagram is such there can be no Kerr effect at $\Delta _{B}=\Delta
_{M}.$

We extend our previous study of the phase diagram\cite%
{ReneSU8,ReneOrbital,ReneHelical} to the case where there is an electric
field applied in a direction parallel to the layers. It was shown before
that such a field would gap the orbital pseudospin Goldstone mode of the
uniform phase with orbital coherence\cite{Shizuya}. We show in this work how
the phase diagram is modified by such a field, how the collective modes
dispersion are changed in the various phases and finally how this field
affects the optical absorption.

This paper is organized as follow. We describe in Sec. II the 2DEG at
filling factor $\nu =3$ as an effective two-level system with frozen valley
and spin degrees of freedom. We derive in Sec. III the phase diagram at $\nu
=3$ when one of the two level is filled. In Sec. IV, we consider the effect
of a parallel electric field on the phase diagram. In Secs. V and VI, we
derive the optical absorption and Kerr effect due to the collective
excitations. We conclude in Sec. VII. To avoid repetitions, we refer the
reader to previous works in Refs. %
\onlinecite{ReneSU8,ReneOrbital,ReneHelical}, where the Hartree-Fock method
for deriving the phase diagram and the calculation of the collective modes
in the generalized random-phase approximation (GRPA) are described in detail.

\section{THE$\ $2DEG$\ $AT $\protect\nu =3$ AS A\ TWO-LEVEL\ SYSTEM\qquad
\qquad}

The system considered in this work is a Bernal-stacked graphene bilayer
(BLG) in a transverse magnetic $\mathbf{B=}B\widehat{\mathbf{z}}$ and
electric $\mathbf{E=}E\widehat{\mathbf{z}}$ fields$\mathbf{.}$ The electric
field induces a potential difference (or \textit{bias}) $\Delta _{B}=Ed$
between the two layers separated by a distance $d=3.34$ \AA . The honeycomb
lattice in each of these layers is described as a triangular Bravais lattice
with a basis of two carbon atoms $A_{n}$ and $B_{n},$ where $n=1,2$ is the
layer index and the lattice constant is $a_{0}=2.\,\allowbreak 46$ \AA . The
unit cell has four lattice sites denoted by $\left\{
A_{1},B_{1},A_{2},B_{2}\right\} .$ The reciprocal lattice has an hexagonal
Brillouin zone with two non-equivalent valleys $K_{\xi }=\left( \frac{2\pi }{%
a_{0}}\right) \left( \xi \frac{2}{3},0\right) ,$ where $\xi =\pm $ $1$.\cite%
{BilayerReview} In the Bernal stacking arrangement, the upper $A$ sublattice
is directly on top of the lower $B$ sublattice while the upper $B$
sublattice is above the center of a hexagonal plaquette of the lower layer.

For a neutral bilayer, the chemical potential is at the energy $E=0$ and the
low-energy excitations ($E<<\gamma _{1}$) can be studied using an effective
two-component model\cite{McCann} with an Hamiltonian, in the absence of the
quantizing magnetic field, given by%
\begin{eqnarray}
&&H_{\xi ,\sigma }^{0}\left( \mathbf{p}\right) =  \label{nonh} \\
&&\left( 
\begin{array}{cc}
\begin{array}{c}
\xi \frac{\Delta _{B}}{2}+\eta _{-\xi }p_{-}p_{+} \\ 
-\frac{1}{2}\sigma \Delta _{Z}%
\end{array}
& \frac{1}{2m^{\ast }}p_{-}^{2} \\ 
\frac{1}{2m^{\ast }}p_{+}^{2} & 
\begin{array}{c}
-\xi \frac{\Delta _{B}}{2}+\eta _{\xi }p_{+}p_{-} \\ 
-\frac{1}{2}\sigma \Delta _{Z}%
\end{array}%
\end{array}%
\right) .  \notag
\end{eqnarray}%
This Hamiltonian is here written in the basis $\left( A_{2},B_{1}\right) $
for valley $K_{-}$ and $\left( B_{1},A_{2}\right) $ for valley $K_{+}$ and $%
p_{\pm }=p_{x}\pm ip_{y}.$ The parameter 
\begin{equation}
\eta _{\xi }=\frac{1}{2m^{\ast }}\left( \xi \frac{\Delta _{B}}{\gamma _{1}}+2%
\frac{\gamma _{4}}{\gamma _{0}}+\frac{\delta _{0}}{\gamma _{1}}\right) ,
\end{equation}%
with the effective mass $m^{\ast }=2\hslash ^{2}\gamma _{1}/3\gamma
_{0}^{2}a_{0}^{2},$ where $\gamma _{0}=2.61$ eV\cite{RecentValues} is the
in-plane nearest-neighbor hopping, $\gamma _{1}=-0.361$ eV is the interlayer
hopping between carbon atoms that are immediately above one another (i.e. $%
A_{1}-B_{2}$) and $\gamma _{4}=-0.138$ eV is the interlayer next
nearest-neighbor hopping term between carbons atoms in the same sublattice
(i.e. $A_{1}-A_{2}$ and $B_{1}-B_{2}$). The parameter $\delta _{0}=0.015$ eV
represents the difference in the crystal field between sites $A_{1},B_{2}$
and $A_{2},B_{1}$. We ignore the warping term $\gamma _{3}$, a valid
approximation at the magnetic fields considered in this article.\cite%
{McCann,ReneValidity} The Zeeman coupling $\Delta _{Z}=g\mu _{B}B,$ where $%
g=2$ and $\sigma =\pm 1$ is the spin index.

A quantizing perpendicular magnetic field is taken into account by making
the Peierls substitution $\mathbf{p}\rightarrow \mathbf{P}=\mathbf{p}+e%
\mathbf{A}/c$ (with $e>0$), where $\nabla \times \mathbf{A=}B\widehat{%
\mathbf{z}}.$ Defining the ladder operators $a=\left( P_{x}-iP_{y}\right)
\ell /\sqrt{2}\hslash $ and $a^{\dag }=\left( P_{x}+iP_{y}\right) \ell /%
\sqrt{2}\hslash $ and the magnetic length $\ell =\sqrt{\hslash c/eB},$ we get%
\begin{equation}
H_{\xi ,\sigma }^{0}=\left( 
\begin{array}{cc}
\begin{array}{c}
\xi \frac{\Delta _{B}}{2}+\zeta _{-}aa^{\dag } \\ 
-\frac{1}{2}\sigma \Delta _{Z}%
\end{array}
& \zeta ^{\prime }a^{2} \\ 
\zeta ^{\prime }\left( a^{\dag }\right) ^{2} & 
\begin{array}{c}
-\xi \frac{\Delta _{B}}{2}+\zeta _{+}a^{\dag }a \\ 
-\frac{1}{2}\sigma \Delta _{Z}%
\end{array}%
\end{array}%
\right) ,  \label{s1}
\end{equation}%
where%
\begin{eqnarray}
\zeta &=&\beta \left( 2\frac{\gamma _{1}\gamma _{4}}{\gamma _{0}}+\delta
_{0}\right) , \\
\zeta _{\pm } &=&\zeta \pm \xi \beta \Delta _{B}, \\
\zeta ^{\prime } &=&\beta \gamma _{1}\left( 1+2\frac{\delta _{0}\gamma _{4}}{%
\gamma _{0}\gamma _{1}}+\left( \frac{\gamma _{4}}{\gamma _{0}}\right)
^{2}\right) ,
\end{eqnarray}%
and%
\begin{equation}
\beta =\frac{\hslash \omega _{c}^{\ast }}{\gamma _{1}}=7.24\times 10^{-3}B%
\text{[T],}
\end{equation}%
where the effective cyclotron frequency $\omega _{c}^{\ast }=eB/m^{\ast }c$
with%
\begin{equation}
\hslash \omega _{c}^{\ast }=2.\,\allowbreak 61B\text{[T] meV.}
\end{equation}
In Eq. (\ref{s1}), the ladder operators are defined by $a^{\dag }\varphi
_{n}\left( x\right) =i\sqrt{n+1}\varphi _{n+1}\left( x\right) $ and $%
a\varphi _{n}\left( x\right) =-i\sqrt{n}\varphi _{n-1}\left( x\right) ,$
where $\varphi _{n}\left( x\right) $ (with $n=0,1,2,...$) are the
eigenfunctions of the one-dimensional harmonic oscillator.

The two-component model describes well\cite{ReneValidity} the eigenenergies
and eigenstates of the Landau level $N=0$ which has eight sub-levels indexed
by the quantum numbers $\xi $ and $\sigma $ and an extra "orbital" index $%
n=0,1$. The eigenenergies are%
\begin{eqnarray}
E_{\xi ,\sigma ,n=0} &=&-\xi \frac{\Delta _{B}}{2}-\sigma \frac{\Delta _{Z}}{%
2},  \label{ener1} \\
E_{\xi ,\sigma ,n=1} &=&-\xi \frac{\Delta _{B}}{2}-\sigma \frac{\Delta _{Z}}{%
2}+\zeta +\xi \beta \Delta _{B}.  \label{ener2}
\end{eqnarray}%
At zero bias, the degeneracy of the octet of states in $N=0$ is lifted by
the Zeeman, the $\gamma _{4}$ and the $\delta _{0}$ terms. These couplings
are small however since%
\begin{eqnarray}
\zeta &=&0.39B\text{[T] meV,} \\
\Delta _{Z} &=&0.12B\text{[T] meV.}
\end{eqnarray}%
In the simplest tight-binding model where $\gamma _{4}=\delta _{0}=0$ and
with $\Delta _{Z}=\Delta _{B}=0,$ the Landau level spectrum is given by $%
E_{N}=sgn\left( N\right) \sqrt{\left\vert N\right\vert \left( \left\vert
N\right\vert +1\right) }\hslash \omega _{c}^{\ast }$ where $N=0,\pm 1,\pm
2,\ldots $and $sgn$ is the signum function. The gap between the first two
Landau levels is $E_{1}-E_{0}\approx 37$ meV. The two-component spinors for
the levels $\left( \xi ,\sigma ,n=0,1\right) $ are independent of the spin
index $\sigma $ and given, in the common basis $\left( A_{2},B_{1}\right) ,$
by 
\begin{eqnarray}
\psi _{\xi =+1,n,X}\left( \mathbf{r}\right) &=&\left( 
\begin{array}{c}
h_{n,X}\left( \mathbf{r}\right) \\ 
0%
\end{array}%
\right) , \\
\psi _{\xi =-1,n,X}\left( \mathbf{r}\right) &=&\left( 
\begin{array}{c}
0 \\ 
h_{n,X}\left( \mathbf{r}\right)%
\end{array}%
\right) ,
\end{eqnarray}%
with the Landau-level wave functions in the Landau gauge $\mathbf{A}=\left(
0,Bx,0\right) $ given by 
\begin{equation}
h_{n,X}\left( \mathbf{r}\right) =\frac{1}{\sqrt{L_{y}}}e^{-iXy/\ell
^{2}}\varphi _{n}\left( x-X\right) ,
\end{equation}%
where $X$ is the guiding-center index.

Crossings between some levels of $N=0$ and $N=1$ occur at bias $\left\vert
\Delta _{B}\right\vert \approx 0.1$ eV for $B=10$ T and $\left\vert \Delta
_{B}\right\vert \approx 0.2$ eV for $B=30$ T. Above these biases, it is in
principle not possible to neglect Landau-level mixing.\cite{ReneValidity}

The $T=0$ K phase diagram of the chiral 2DEG in $N=0$ has been derived in
Ref. \onlinecite{ReneSU8} in the two-component model and in the Hartree-Fock
approximation for integer filling factors $\nu \in \left[ -3,3\right] $ as a
function of the bias $\Delta _{B}$. In the present work, we study in more
detail the phase diagram for filling factor $\nu =3$ which corresponds to
the filling of seven sub-levels in $N=0.$ According to Ref. %
\onlinecite{ReneSU8}, this means that all four spin up states are filled as
well as the two states $n=0,1$ with $\sigma =-1$ in valley $K_{+}$.
Considering all these states as inert, we are left with a simple two-level
system consisting of the two orbital states $n=0,1$ in valley $K_{-}$ with $%
\sigma =-1.$ We denote by $\widetilde{\nu }$ the filling factor of this
two-level system so that $\widetilde{\nu }=1$ when $\nu =3.$ The order
parameters of the different phases are given by%
\begin{eqnarray}
\left\langle \rho _{n,m}\left( \mathbf{q}\right) \right\rangle &=&\frac{1}{%
N_{\varphi }}\sum_{X,X^{\prime }}e^{-\frac{i}{2}q_{x}\left( X+X^{\prime
}\right) }  \label{order2} \\
&&\times \left\langle c_{X,n}^{\dagger }c_{X^{\prime },m}\right\rangle
\delta _{X,X^{\prime }+q_{y}\ell ^{2}},  \notag
\end{eqnarray}%
where $N_{\varphi }$ is the Landau-level degeneracy and $\mathbf{q}$\textbf{%
\ }is a two-dimensional vector in the plane of the 2DEG. Hereafter, we drop
the indices $\xi =-1,\sigma =-1$ to simplify the notation. The diagonal
elements $\left\langle \rho _{n,n}\left( \mathbf{q}=0\right) \right\rangle =%
\widetilde{\nu }_{n}$ give the filling factor of each level $n$ while the
off-diagonal elements describe orbital coherence.

These average electronic density $n\left( \mathbf{q}\right) $ is given by%
\begin{equation}
n\left( \mathbf{q}\right) =N_{\varphi }\sum_{n,m}K_{n,m}\left( -\mathbf{q}%
\right) \left\langle \rho _{n,m}\left( \mathbf{q}\right) \right\rangle ,
\label{densityp}
\end{equation}%
where the form factors 
\begin{eqnarray}
K_{0,0}\left( \mathbf{q}\right) &=&e^{\frac{-q^{2}\ell ^{2}}{4}}, \\
K_{1,1}\left( \mathbf{q}\right) &=&e^{\frac{-q^{2}\ell ^{2}}{4}}\left( 1-%
\frac{q^{2}\ell ^{2}}{2}\right) , \\
K_{1,0}\left( \mathbf{q}\right) &=&e^{\frac{-q^{2}\ell ^{2}}{4}}\left( \frac{%
\left( q_{y}+iq_{x}\right) \ell }{\sqrt{2}}\right) , \\
K_{0,1}\left( \mathbf{q}\right) &=&e^{\frac{-q^{2}\ell ^{2}}{4}}\left( \frac{%
\left( -q_{y}+iq_{x}\right) \ell }{\sqrt{2}}\right) ,
\end{eqnarray}%
capture the orbital character of the two states.

In a pseudospin language, the states $n=0,1$ are represented by the up and
down pseudospin states respectively. The orbital pseudospin vector $\mathbf{p%
}\left( \mathbf{q}\right) $ is related to the operators $\rho _{n,m}\left( 
\mathbf{q}\right) $ by%
\begin{eqnarray}
\rho \left( \mathbf{q}\right) &=&\rho _{0,0}\left( \mathbf{q}\right) +\rho
_{1,1}\left( \mathbf{q}\right) , \\
p_{x}\left( \mathbf{q}\right) &=&\left[ \rho _{0,1}\left( \mathbf{q}\right)
+\rho _{1,0}\left( \mathbf{q}\right) \right] /2, \\
p_{y}\left( \mathbf{q}\right) &=&\left[ \rho _{0,1}\left( \mathbf{q}\right)
-\rho _{1,0}\left( \mathbf{q}\right) \right] /2i, \\
p_{z}\left( \mathbf{q}\right) &=&\left[ \rho _{0,0}\left( \mathbf{q}\right)
-\rho _{1,1}\left( \mathbf{q}\right) \right] /2,
\end{eqnarray}%
and $\left\vert \left\langle \mathbf{p}\left( 0\right) \right\rangle
\right\vert =1/2$ for $\widetilde{\nu }=1.$

The phase diagram at $\widetilde{\nu }=1$ contains both uniform and
nonuniform states with pseudospin textures. The Hartree-Fock energy for a
state with a pseudospin texture is given by\cite{ReneHelical} (apart from
some unimportant constant terms) by

\begin{eqnarray}
&&\frac{E_{HF}}{N}=-\beta \left( \Delta _{M}-\Delta _{B}\right) \left\langle 
\widetilde{p}_{z}\left( 0\right) \right\rangle  \label{texture} \\
&&+\frac{1}{2}\left( \frac{e^{2}}{\kappa \ell }\right) \sum_{\mathbf{q}%
}\left\langle \widetilde{\mathbf{p}}_{\Vert }\left( \mathbf{-q}\right)
\right\rangle \cdot \left[ a\left( q\right) \mathbf{I}+b\left( q\right) 
\mathbf{\Lambda }\left( \mathbf{q}\right) \right] \cdot \left\langle 
\widetilde{\mathbf{p}}_{\Vert }\left( \mathbf{q}\right) \right\rangle  \notag
\\
&&+\frac{1}{2}\left( \frac{e^{2}}{\kappa \ell }\right) \sum_{\mathbf{q}%
}c\left( q\right) \left\langle \widetilde{p}_{z}\left( -\mathbf{q}\right)
\right\rangle \left\langle \widetilde{p}_{z}\left( \mathbf{q}\right)
\right\rangle  \notag \\
&&+\frac{i}{4}\left( \frac{e^{2}}{\kappa \ell }\right) \sum_{\mathbf{q}%
}d\left( q\right) \left( \widehat{\mathbf{z}}\times \widehat{\mathbf{q}}%
\right) \cdot \left[ \left\langle \widetilde{\mathbf{p}}\left( \mathbf{-q}%
\right) \right\rangle \times \left\langle \widetilde{\mathbf{p}}\left( 
\mathbf{q}\right) \right\rangle \right] ,  \notag
\end{eqnarray}%
where $\widetilde{\mathbf{p}}\equiv \left( -p_{x},p_{y},p_{z}\right) ,$ $%
\mathbf{I}$ is the $2\times 2$ unit tensor and 
\begin{equation}
\mathbf{\Lambda }\left( \mathbf{q}\right) =\left( 
\begin{array}{cc}
\cos \left( 2\varphi _{\mathbf{q}}\right) & \sin \left( 2\varphi _{\mathbf{q}%
}\right) \\ 
\sin \left( 2\varphi _{\mathbf{q}}\right) & -\cos \left( 2\varphi _{\mathbf{q%
}}\right)%
\end{array}%
\right) ,
\end{equation}%
where $\varphi _{\mathbf{q}}$ is the angle between the wave vector $\mathbf{q%
}$ and the $x$ axis. The interactions $a\left( q\right) ,b\left( q\right)
,c\left( q\right) $ and $d\left( q\right) $ are defined in Ref. %
\onlinecite{ReneHelical}. The bias $\Delta _{M}$ defines the middle of the
spiral phase in Fig. 1.

The interaction energy of the 2DEG with a uniform external electric field $%
\mathbf{E}_{\Vert }=-\nabla V\left( \mathbf{r}\right) $ applied in the plane
of the bilayer is given by

\begin{eqnarray}
\left\langle H_{\mathbf{E}_{\Vert }}\right\rangle &=&-e\int d\mathbf{r}%
n\left( \mathbf{r}\right) V\left( \mathbf{r}\right)  \label{dipole} \\
&=&-\int d\mathbf{rd}\left( \mathbf{r}\right) \cdot \mathbf{E}_{\Vert }, 
\notag
\end{eqnarray}%
with the average density $n\left( \mathbf{r}\right) $ defined in Eq. (\ref%
{densityp}) and the average total electric dipole $\mathbf{d}\left( \mathbf{q%
}\right) $ related to the pseudospin operator $\widetilde{\mathbf{p}}$ by%
\cite{Shizuya} 
\begin{equation}
\mathbf{d}\left( \mathbf{q}\right) =\sqrt{2}e\ell N_{\varphi }e^{\frac{%
-q^{2}\ell ^{2}}{4}}\left\langle \widetilde{\mathbf{p}}_{\Vert }\left( 
\mathbf{q}\right) \right\rangle .  \label{coherence}
\end{equation}%
To get Eq. (\ref{dipole}), we have taken into account that, for all phases
studied in this work, the condition $\left\langle \rho \left( \mathbf{q}%
\right) \right\rangle =\delta _{\mathbf{q},0}$ is satisfied.

\section{PHASE\ DIAGRAM FOR\ $\protect\widetilde{\protect\nu }=1$}

For $\mathbf{E}_{\Vert }=0,$ the sequence of ground states\cite%
{ReneOrbital,ReneHelical,ReneSU8} as the bias is increased is illustrated in
Fig. 1. We use the values $B=10$ T and $\kappa =5$ for the host dielectric
constant in all our numerical calculations. The different phases are as
follows:

%%%%%%%%%%%%%%%%%%%%%%%%%%%%%%%%%%%%%%%%%%%%%%%%%%%%%%%%%%%%%%%%%%%%%%%%%%
\begin{figure}[tbph]
\includegraphics[scale=1.0]{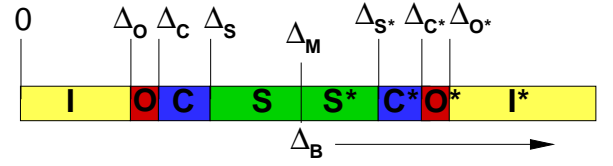}
\caption{(Color online) Phase diagram of the 2DEG in bilayer graphene at
filling factor $\protect\widetilde{\protect\nu }=1$. The biases $\Delta _{i}$
indicate the onset of each phase. (The bias axis is not to scale.)}
\label{figure1}
\end{figure}
%%%%%%%%%%%%%%%%%%%%%%%%%%%%%%%%%%%%%%%%%%%%%%%%%%%%%%%%%%%%%%%%%%%%%%%%%%

\begin{itemize}
\item Phase $I$ for $\Delta _{B}\in \left[ 0,\Delta _{O}=\zeta /\beta \right]
$ is a uniform phase where all pseudospins are oriented along $\widehat{%
\mathbf{z}}$ (i.e. $\widetilde{\nu }_{0}=1$)$.$ The Hartree-Fock gap (the
energy to create an electron-hole pair) is given by $\Delta _{eh}=\eta
/2+\beta \left( \Delta _{O}-\Delta _{B}\right) $ and is shown in Fig. 5.
Phase $I$ has a pseudospin-wave mode with a dispersion $\omega _{I}\left( 
\mathbf{q}\right) ,$ shown in Fig. 4(a), with a gap $\omega _{I}\left(
0\right) =\beta \left( \Delta _{O}-\Delta _{B}\right) /\hslash $. This phase
becomes unstable at $\Delta _{O}=\zeta /\beta $ which is the onset of the
coherent phase $O.$ The energy%
\begin{equation}
\eta =\sqrt{\frac{\pi }{2}}\left( \frac{e^{2}}{\kappa \ell }\right) =\frac{%
56.1\sqrt{B\text{[T]}}}{\kappa }\text{meV}.
\end{equation}

\item Phase $O$ for $\Delta _{B}\in \left[ \Delta _{O},\Delta _{C}\right] $
is a uniform coherent phase where all the pseudospins are tilted by an angle 
$\theta $ with respect to the $z$ axis with $\cos \left( \theta \right)
=1-8\beta \left( \Delta _{B}-\Delta _{O}\right) /\eta .$ The Hartree-Fock
gap $\Delta _{eh}=\eta /2$ is constant in this phase as shown by the dashed
line in Fig. 5. The energy of this phase is invariant with respect to a
collective rotation of the pseudospins around the $z$ axis. It thus has a
gapless pseudospin-wave mode whose dispersion, shown in Fig. 4(b), is highly
anisotropic in the $x-y$ plane.\cite{ReneOrbital} Figure 4(b) shows the
dispersion in the directions parallel ($x$) and perpendicular ($y$) to the
dipoles when they are oriented in the $\widehat{\mathbf{x}}$ direction. The
origin of the orbital coherence is easy to understand. From Eqs. (\ref{ener1}%
)-(\ref{ener2}), $E_{n=1}>E_{n=0}$ at $\Delta _{B}=0$ since $\zeta >0.$ As
the bias is increased, however, $E_{n=1}\rightarrow E_{n=0}$ until $%
E_{n=1}\leq E_{n=0}$ above a critical bias $\Delta _{B}=\zeta /\beta .$ The
electrons should then occupy state $n=1$ instead of $n=0.$ But, the Coulomb
exchange energy is smaller when the electrons occupy level $n=0$ instead of $%
n=1.$ To resolve this conflict, the system optimizes its energy by
delocalizing the electrons in the two levels i.e. by creating a coherent
state. When the small parameters $\gamma _{4},\delta _{0}$ are neglected,
this coherence occurs at zero bias. The orbital coherence is associated with
a finite electric polarization\cite{Shizuya} in the $x-y$ plane as shown in
Eq. (\ref{coherence}). Fluctuations of these dipoles are responsible for the
electromagnetic absorption.

\item Phase $C$ for $\Delta _{B}\in \left[ \Delta _{C},\Delta _{S}\right] $
is a triangular Wigner crystal state with one electron per site and a
vortex-like texture of pseudospins around each site as illustrated in Fig.
2. This texture is described by the Fourier components $\left\langle \mathbf{%
p}\left( \mathbf{G}\right) \right\rangle ,$where $\left\{ \mathbf{G}\right\} 
$ is the set of reciprocal lattice vectors of the crystal. The crystal has a
gapless phonon mode, which has the characteristic long-wavelength dispersion 
$\omega \sim \left( q\ell \right) ^{1.5}$ of a Wigner crystal in a magnetic
field. There are also higher-energy gapped modes whose dispersions are shown
in Fig. 4(c) and an electron-hole continuum of excitations.\cite{ReneHelical}
The Hartree-Fock gap $\Delta _{eh}$ for this phase is shown in Fig. 5.

\item Phase $S$ for $\Delta _{B}\in \left[ \Delta _{S},\Delta _{S^{\ast }}%
\right] $ is shown in Fig. 3. It is a spiral phase where the pseudospins
rotate in the plane $z-n$ plane with $\widehat{\mathbf{n}}$ being some
arbitrary direction in the $x-y$ plane. The Fourier components $\left\langle 
\mathbf{p}\left( mQ\widehat{\mathbf{n}}\right) \right\rangle \neq 0,$ where $%
m=0,\pm 1,\pm 2,...$ and $Q$ is the wave vector of the spiral. The energy of
the spiral is independent of the orientation of $\widehat{\mathbf{n}}$ and
so it has a gapless phonon mode. There are also higher-energy gapped mode
whose dispersion are shown in Fig. 4(d) and an electron-hole continuum of
excitations.\cite{ReneHelical} The dispersion of these modes is highly
anisotropic. The gap $\Delta _{eh}$ for this phase is shown in Fig. 5.
\end{itemize}

The phase diagram is symmetrical with respect to the bias $\Delta _{M}=\zeta
/\beta +\eta /8\beta $ which is in the middle of the spiral phase. We say
that a phase with $\Delta _{B}^{\left( 2\right) }=\Delta _{M}+\Delta $ is
the \textit{conjugate} of that with $\Delta _{B}^{\left( 1\right) }=\Delta
_{M}-\Delta $ (with $\Delta >0$) in the sense that it has the same gap $%
\Delta _{eh}$ and the same spectrum of collective modes. The filling factors 
$\widetilde{\nu }_{0}>\widetilde{\nu }_{1}$ for $\Delta _{B}^{\left(
1\right) }$ and vice versa for $\Delta _{B}^{\left( 2\right) }.$ We denote
the conjugate phases with $\Delta _{B}>\Delta _{M}$ by $S^{\ast },C^{\ast
},O^{\ast },I^{\ast }.$ In phase $I^{\ast },$ the pseudospin are aligned
along $-\widehat{\mathbf{z}}$ (i.e. $\widetilde{\nu }_{1}=1$), the gap of
the pseudospin-wave mode is given by $\omega _{I^{\ast }}\left( 0\right)
=\beta \left( \Delta _{B}-\Delta _{O^{\ast }}\right) /\hslash $ and $\Delta
_{eh}=\eta /2+\beta \left( \Delta _{B}-\Delta _{O^{\ast }}\right) .$ The
bias $\Delta _{O^{\ast }}=\zeta /\beta +\eta /4\beta .$ There are no more
phase transitions for $\Delta _{B}>\Delta _{O^{\ast }}$ in the two-level
system.

%%%%%%%%%%%%%%%%%%%%%%%%%%%%%%%%%%%%%%%%%%%%%%%%%%%%%%%%%%%%%%%%%%%%%%%%%%
\begin{figure}[tbph]
\includegraphics[scale=0.8]{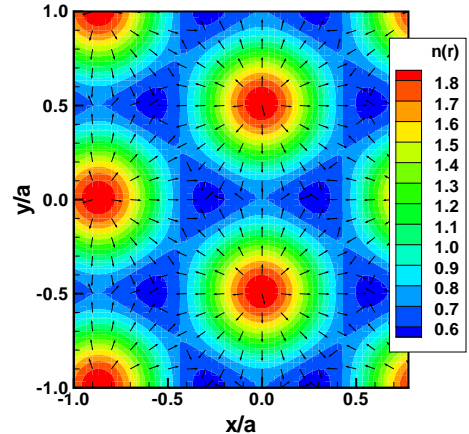}
\caption{(Color online) Electronic
density $n\left( \mathbf{r}\right) $ and dipole pattern (arrows) $\mathbf{d}\left( 
\mathbf{r}\right) $ for the Wigner crystal phase at $\Delta _{B}=$ $59$ meV.}
\label{figure2}
\end{figure}
%%%%%%%%%%%%%%%%%%%%%%%%%%%%%%%%%%%%%%%%%%%%%%%%%%%%%%%%%%%%%%%%%%%%%%%%%%

%%%%%%%%%%%%%%%%%%%%%%%%%%%%%%%%%%%%%%%%%%%%%%%%%%%%%%%%%%%%%%%%%%%%%%%%%%
\begin{figure}[tbph]
\includegraphics[scale=0.8]{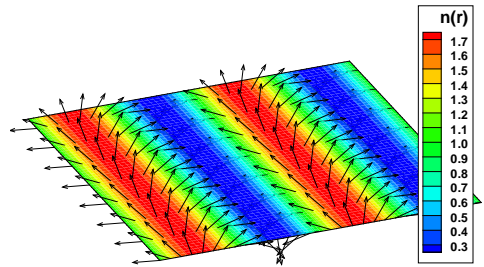}
\caption{(Color online) Electronic density 
$n\left( \mathbf{r}\right) $. The arrows represent the dipole $\mathbf{d}\left( \mathbf{r%
}\right) $ and pseudospin fields $\left\langle p_{z}\left( \mathbf{r}%
\right) \right\rangle $ of the spiral phase for $\Delta _{B}=112$ meV.}
\label{figure3}
\end{figure}
%%%%%%%%%%%%%%%%%%%%%%%%%%%%%%%%%%%%%%%%%%%%%%%%%%%%%%%%%%%%%%%%%%%%%%%%%%

%%%%%%%%%%%%%%%%%%%%%%%%%%%%%%%%%%%%%%%%%%%%%%%%%%%%%%%%%%%%%%%%%%%%%%%%%%
\begin{figure}[tbph]
\includegraphics[scale=1.0]{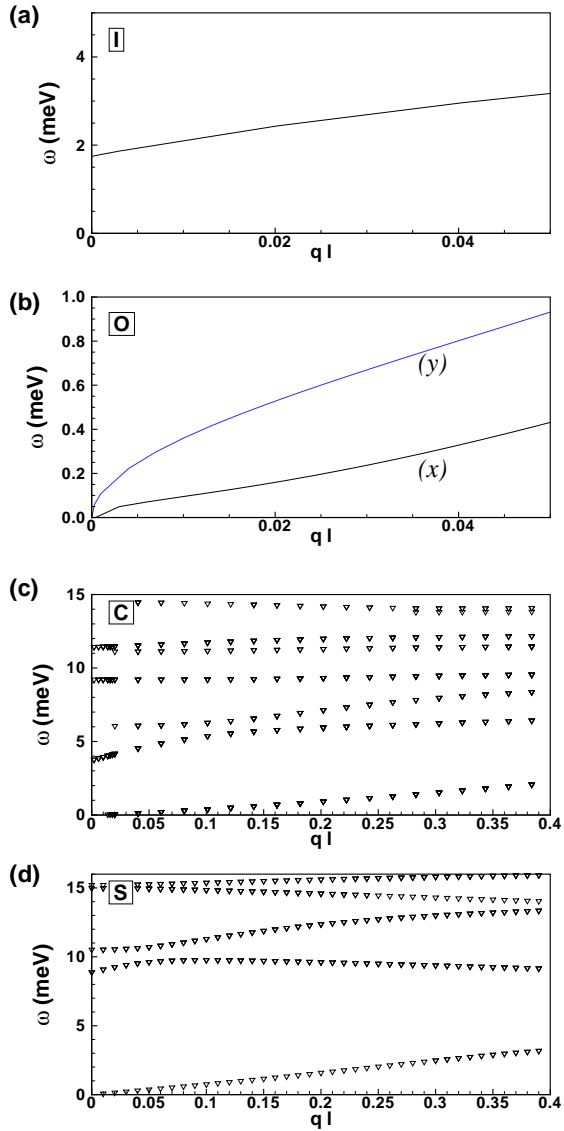}
\caption{(Color online) Dispersion
relations for the (a) incoherent phase $I$ at $\Delta _{B}=29$ meV; (b)
coherent phase $O$ at $\Delta _{B}=54$ meV; (c) crystal phase $C$ at $\Delta
_{B}=66$ meV, and (d) spiral phase $S$ at $\Delta _{B}=77$ meV. }
\label{figure4}
\end{figure}
%%%%%%%%%%%%%%%%%%%%%%%%%%%%%%%%%%%%%%%%%%%%%%%%%%%%%%%%%%%%%%%%%%%%%%%%%%

%%%%%%%%%%%%%%%%%%%%%%%%%%%%%%%%%%%%%%%%%%%%%%%%%%%%%%%%%%%%%%%%%%%%%%%%%%
\begin{figure}[tbph]
\includegraphics[scale=1.0]{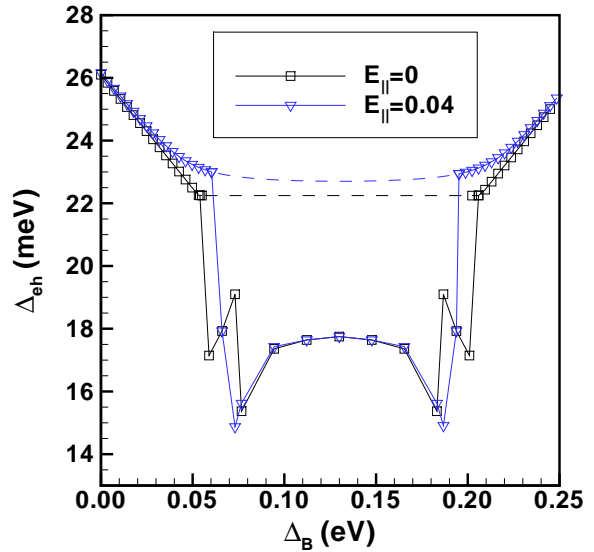}
\caption{(Color online) Hartree-Fock electron-hole gap $\Delta _{eh}$ as a
function of bias for two values of the parallel electric field $\mathbf{E}%
_{\Vert }=E_{\Vert }\protect\widehat{\mathbf{x}}$ in mV/nm. The dashed lines
indicate the gap in the uniform coherent phase $O$ when it is not the ground
state. This phase is replaced by the crystal and spiral phases with smaller
gap in most of the phase diagram.}
\label{figure5}
\end{figure}
%%%%%%%%%%%%%%%%%%%%%%%%%%%%%%%%%%%%%%%%%%%%%%%%%%%%%%%%%%%%%%%%%%%%%%%%%%

With the hopping parameters given in Sec. II, the critical biases in meV
are: $\Delta _{O}=53.2$, $\Delta _{C}=55.7,\Delta _{S}=75.6,\Delta _{S^{\ast
}}=184.6,\Delta _{M}=130.1,\Delta _{C^{\ast }}=204.5,\Delta _{O^{\ast
}}=207.0.$ Thus, at $B=10$ T and $\kappa =5$, only the first half of the
phase diagram falls below the limits of validity of the two-component model.
But, the full phase diagram should be visible at higher magnetic field. At $%
B=30$ T, for example, $\Delta _{O^{\ast }}<200$ meV which is within the
limits of validity of the model.\cite{ReneValidity}

\section{EFFECT OF A PARALLEL ELECTRIC FIELD ON THE PHASE DIAGRAM\qquad}

Adding the dipole term of Eq. (\ref{dipole}) to the Hartree-Fock Hamiltonian
and using the formalism described in Refs. \onlinecite{ReneSU8,ReneHelical}
to compute the single-particle Green's function, we obtain the phase diagram
shown in Fig. 6. With a finite parallel electric field $\mathbf{E}_{\Vert },$
the pseudospins are pushed towards the $x-y$ plane. The $I$ and $I^{\ast }$
phases are transformed into the $O$ and $O^{\ast }$ phases and orbital
coherence is then always present. We find that the orientation of the wave
vector $\mathbf{Q}$ that minimizes the Hartree-Fock energy is $Q\bot \mathbf{%
E}_{\Vert }.$ From Fig. 6, it can be seen that, although a small electric
field is sufficient to suppress the crystal phase, a much larger field is
needed to destroy the spiral phase. This field is 0.30 mV/nm at $\Delta_{B}=0.13$ eV 
in the middle of the spiral phase (not shown in the figure). The 2DEG is described by the pseudospin
energy functional given in Eq. (\ref{texture}) where the effective
pseudospin Heisenberg exchange interaction is highly anisotropic. Figure 7
gives an idea of the strength of the parallel electric field which is
necessary to tilt the pseudospin away from the $z$ axis, against the
interaction $c(q)<0$ that tends to keep them aligned with that axis.

%%%%%%%%%%%%%%%%%%%%%%%%%%%%%%%%%%%%%%%%%%%%%%%%%%%%%%%%%%%%%%%%%%%%%%%%%%
\begin{figure}[tbph]
\includegraphics[scale=1.0]{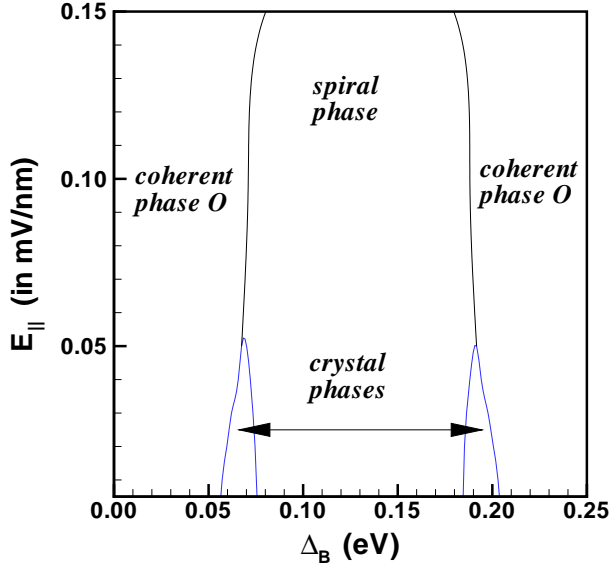}
\caption{(Color online) Phase diagram
of the 2DEG as a function of the bias $\Delta _{B}$ and the electric field $%
\mathbf{E}_{\Vert }$. }
\label{figure6}
\end{figure}
%%%%%%%%%%%%%%%%%%%%%%%%%%%%%%%%%%%%%%%%%%%%%%%%%%%%%%%%%%%%%%%%%%%%%%%%%%

%%%%%%%%%%%%%%%%%%%%%%%%%%%%%%%%%%%%%%%%%%%%%%%%%%%%%%%%%%%%%%%%%%%%%%%%%%
\begin{figure}[tbph]
\includegraphics[scale=1.0]{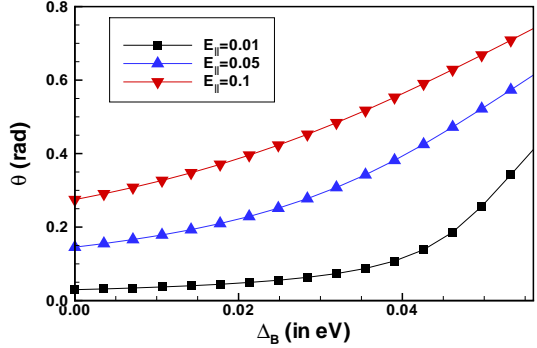}
\caption{(Color online) In phase $I,$ a
parallel electric field $\mathbf{E}_{\Vert }$ tilts the pseudospins towards
the $x-y$ plane by an angle $\protect\theta \left( \Delta _{B},E_{\Vert
}\right) .$ The field $E_{\Vert }$ in the legend is in mV/nm.  }
\label{figure7}
\end{figure}
%%%%%%%%%%%%%%%%%%%%%%%%%%%%%%%%%%%%%%%%%%%%%%%%%%%%%%%%%%%%%%%%%%%%%%%%%%

When $E_{\Vert }\neq 0,$ the Hartree-Fock electron-hole gap is modified in
the manner shown in Fig. 5. The gap is only slightly increased in the
incoherent and coherent phases and does not change noticeably in the crystal
and spiral phases for an electric field $E_{\Vert }=0.04$ mV/nm which is
near the upper-limit of the crystal phase in Fig. 6.

%%%%%%%%%%%%%%%%%%%%%%%%%%%%%%%%%%%%%%%%%%%%%%%%%%%%%%%%%%%%%%%%%%%%%%%%%%
\begin{figure}[tbph]
\includegraphics[scale=1.0]{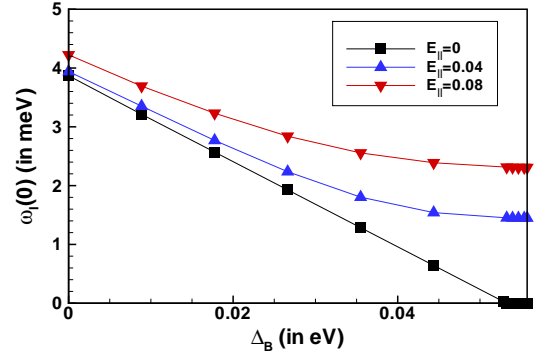}
\caption{(Color online) Optical gaps $%
\protect\omega _{I}\left( 0\right) $ and $\protect\omega _{O}\left( 0\right) 
$ as a function of bias for different values of the parallel electric field
(in mV/nm)$.$ The flat region of each curve corresponds to the coherent phase $O$. }
\label{figure8}
\end{figure}
%%%%%%%%%%%%%%%%%%%%%%%%%%%%%%%%%%%%%%%%%%%%%%%%%%%%%%%%%%%%%%%%%%%%%%%%%%

Figure 8 shows how the pseudospin-wave optical gaps $\omega _{I}\left(
0\right) $ and $\omega _{O}\left( 0\right) $ of the $I$ and $O$ phases
change with bias for different values of the electric field. The electric
field increases the optical gap in the incoherent phase $I.$ In the coherent
phase $O,$ it destroys the $U\left( 1\right) $ symmetry of the Hamiltonian
thus gapping the Goldstone mode.\cite{Shizuya} The gap is practically
constant in the small range of the $O$ phase (the flat region of each curve
in Fig. 5). Note that the optical gap $\omega _{I}\left( 0\right)
\rightarrow 0$ as $\Delta _{B}\rightarrow $ $\Delta _{O}$ when $E_{\Vert
}=0. $ As discussed below, electromagnetic absorption is expected at the gap
frequency.

The dispersion at small wave vector in the crystal phase does not change
very much for an electric field $E_{\Vert }=0.04$ mV/nm. In the spiral
phase, however, the field can be increased to a larger value and it is
possible to modify noticeably the dispersion. In Fig. 9, $E_{\Vert }=0$ and $%
E_{\Vert }=0.15$ mV/nm and the dispersions are shown in the direction of the
spiral. In both the crystal and spiral phases, the phonon mode is not gapped
by a finite $\mathbf{E}_{\Vert }.$ This is easily understood since the
coupling to the external $\mathbf{E}_{\Vert }$ involves the total electric
dipole moment, a quantity that is not changed by a rigid translation of the
system.

%%%%%%%%%%%%%%%%%%%%%%%%%%%%%%%%%%%%%%%%%%%%%%%%%%%%%%%%%%%%%%%%%%%%%%%%%%
\begin{figure}[tbph]
\includegraphics[scale=1.0]{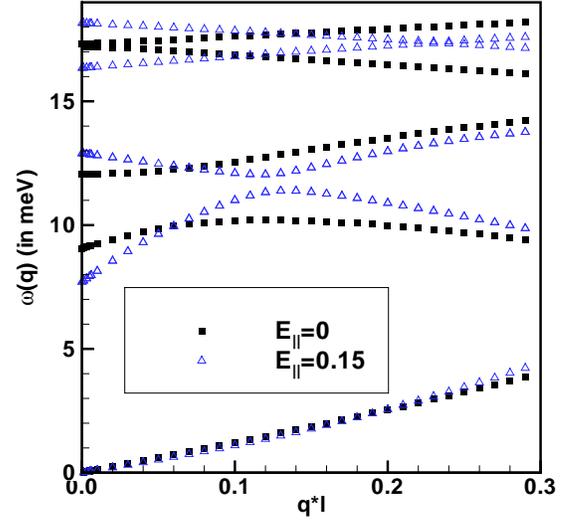}
\caption{(Color online) Dispersion
relation of the collective modes in the direction of the spiral at bias $%
\Delta _{B}=130$ meV for electric field strengths $E_{\Vert }=0$ and $%
E_{\Vert }=0.15$ mV/nm$.$ }
\label{figure9}
\end{figure}
%%%%%%%%%%%%%%%%%%%%%%%%%%%%%%%%%%%%%%%%%%%%%%%%%%%%%%%%%%%%%%%%%%%%%%%%%%

\section{ELECTROMAGNETIC ABSORPTION FROM COLLECTIVE MODES}

In this section, we give detailed derivation of the power absorbed $P\left(
\omega \right) $ by the collective excitations in both the uniform and
nonuniform phases following in parts Ref. \onlinecite{Filund} where $P\left(
\omega \right) $ was calculated for collective modes in quantum wires. The
average power per unit area absorbed from an electromagnetic wave $\mathbf{E}%
_{e}\left( \mathbf{r},z,t\right) $ by a sample of size $S$ located in the $%
x-y$ plane at $z=0$ is given by

\begin{equation}
P\left( \omega \right) =\frac{1}{2S^{2}}\sum_{\mathbf{q}}\Re\left[
\left( \mathbf{E}\left( \mathbf{q},z=0,\omega \right) \right) ^{\ast }\cdot 
\mathbf{j}\left( \mathbf{q},\omega \right) \right] ,
\end{equation}%
where $\mathbf{j}$ is the surface current density in the sample and $\mathbf{%
E}$ is the \textit{total} electric field within the sample. Since nonuniform
as well as uniform phases must be considered, it is necessary to use the
following general relation between the current density and the total
electric field 
\begin{equation}
\mathbf{j}\left( \mathbf{q},\omega \right) =\sum_{\mathbf{q}^{\prime }}%
\overleftrightarrow{\widetilde{\sigma }}\left( \mathbf{q},\mathbf{q}^{\prime
},\omega \right) \cdot \mathbf{E}\left( \mathbf{q}^{\prime },z=0,\omega
\right) ,  \label{cond1}
\end{equation}%
where $\overleftrightarrow{\widetilde{\sigma }}$ is related to the
irreducible or proper part of the two-dimensional current response function.
Equation (\ref{cond1}) can also be written as%
\begin{equation}
\mathbf{j}\left( \mathbf{q},\omega \right) =\sum_{\mathbf{q}^{\prime }}%
\overleftrightarrow{\sigma }\left( \mathbf{q},\mathbf{q}^{\prime },\omega
\right) \cdot \mathbf{E}_{e}\left( \mathbf{q}^{\prime },z=0,\omega \right) ,
\label{cond2}
\end{equation}%
where $\overleftrightarrow{\sigma }$ is now related to the full (i.e.
screened) two-dimensional current response function.\ In our case, it is the
current response computed in the generalized random-phase approximation
(GRPA). In the above relations, $\mathbf{q},\mathbf{q}^{\prime }$ and $%
\mathbf{r}$ are two-dimensional vectors in the plane of the 2DEG. The
absorbed power is thus

\begin{eqnarray}
P\left( \omega \right) &=&\frac{1}{2S^{2}}\sum_{\mathbf{q,q}^{\prime }}\Re%
\left[ \mathbf{E}\left( \mathbf{q},z=0,\omega \right) ^{\ast }\right. \\
&&\cdot \left. \overleftrightarrow{\sigma }\left( \mathbf{q},\mathbf{q}%
^{\prime },\omega \right) \cdot \mathbf{E}_{e}\left( \mathbf{q}^{\prime
},z=0,\omega \right) \right] .  \notag
\end{eqnarray}

From Maxwell equations, the total and external electric fields $\mathbf{E}$
and $\mathbf{E}_{e}$ are related by

\begin{eqnarray}
\mathbf{E}\left( \mathbf{q},z,\omega \right) &=&\mathbf{E}_{e}\left( \mathbf{%
q},z,\omega \right) -\frac{4\pi i}{\omega }\frac{1}{L}\sum_{q_{z}}%
\overleftrightarrow{K}\left( \mathbf{p},\omega \right)  \label{champe} \\
&&\cdot e^{iq_{z}z}\sum_{\mathbf{q}^{\prime }}\overleftrightarrow{\sigma }%
\left( \mathbf{q},\mathbf{q}^{\prime },\omega \right) \cdot \mathbf{E}%
_{e}\left( \mathbf{q}^{\prime },z=0,\omega \right) ,  \notag
\end{eqnarray}%
where the tensor%
\begin{equation}
\overleftrightarrow{K}\left( \mathbf{p},\omega \right) =\widehat{\mathbf{p}}%
\widehat{\mathbf{p}}+\frac{\omega ^{2}}{\omega ^{2}-c^{2}p^{2}}\left( 
\overleftrightarrow{1}-\widehat{\mathbf{p}}\widehat{\mathbf{p}}\right) ,
\end{equation}%
with $\mathbf{p=}\left( \mathbf{q},q_{z}\right) $ and $\overleftrightarrow{1}
$ the three-dimensional unit tensor.

\begin{widetext}

The total electric field is given by

\begin{eqnarray}
\mathbf{E}\left( \mathbf{q},z,\omega \right) &=&\mathbf{E}_{e}\left( \mathbf{%
q},z,\omega \right)  \label{champee} \\
&&-\frac{4\pi i}{\omega }\frac{1}{L}\sum_{q_{z}}\frac{\omega ^{2}e^{iq_{z}z}%
}{\omega ^{2}-c^{2}p^{2}}\sum_{\mathbf{q}^{\prime }}\overleftrightarrow{%
\sigma }\left( \mathbf{q},\mathbf{q}^{\prime },\omega \right) \cdot \mathbf{E%
}_{e}\left( \mathbf{q}^{\prime },z=0,\omega \right)  \notag \\
&&+\frac{4\pi i}{\omega }\frac{1}{L}\sum_{q_{z}}\frac{c^{2}e^{iq_{z}z}}{%
\omega ^{2}-c^{2}p^{2}}\mathbf{pp}\cdot \sum_{\mathbf{q}^{\prime }}%
\overleftrightarrow{\sigma }\left( \mathbf{q},\mathbf{q}^{\prime },\omega
\right) \cdot \mathbf{E}_{e}\left( \mathbf{q}^{\prime },z=0,\omega \right) 
\notag
\end{eqnarray}%
and the absorption by%
\begin{eqnarray}
P\left( \omega \right) &=&\frac{1}{2S^{2}}\sum_{\mathbf{q,q}^{\prime }}\Re%
\left[ \mathbf{E}_{e}^{\ast }\left( \mathbf{q},z=0,\omega \right) \cdot 
\overleftrightarrow{\sigma }\left( \mathbf{q},\mathbf{q}^{\prime },\omega
\right) \cdot \mathbf{E}_{e}\left( \mathbf{q}^{\prime },z=0,\omega \right) %
\right] \\
&&-\frac{1}{2S^{2}}\sum_{\mathbf{q}}\Re\left[ \frac{4\pi i}{\omega }%
\frac{1}{L}\sum_{q_{z}}\frac{\omega ^{2}}{\omega ^{2}-c^{2}p^{2}}\right]
\left\vert \sum_{\mathbf{q}^{\prime }}\overleftrightarrow{\sigma }\left( 
\mathbf{q},\mathbf{q}^{\prime },\omega \right) \cdot \mathbf{E}_{e}\left( 
\mathbf{q}^{\prime },z=0,\omega \right) \right\vert ^{2}  \notag \\
&&+\frac{1}{2S^{2}}\sum_{\mathbf{q}}\Re\left[ \frac{4\pi i}{\omega }%
\frac{1}{L}\sum_{q_{z}}\frac{c^{2}}{\omega ^{2}-c^{2}p^{2}}\right]
\left\vert \sum_{\mathbf{q}^{\prime }}\mathbf{p}\cdot \overleftrightarrow{%
\sigma }\left( \mathbf{q},\mathbf{q}^{\prime },\omega \right) \cdot \mathbf{E%
}_{e}\left( \mathbf{q}^{\prime },z=0,\omega \right) \right\vert ^{2},  \notag
\end{eqnarray}%
where $\omega $ is short for $\omega +i\delta .$

In a crystal, translational symmetry imposes $\overleftrightarrow{\sigma }%
\left( \mathbf{q},\mathbf{q}^{\prime },\omega \right) \rightarrow 
\overleftrightarrow{\sigma }\left( \mathbf{k+G},\mathbf{k+G}^{\prime
},\omega \right) ,$ where $\mathbf{k}$ is a vector in the first Brillouin
zone of the reciprocal lattice and $\mathbf{G},\mathbf{G}^{\prime }$ are
reciprocal lattice vectors. Thus, considering a plane electromagnetic wave $%
\mathbf{E}_{e}\left( \mathbf{q},z=0,\omega \right) =SE_{e}\widehat{\mathbf{e}%
}_{p}\delta _{\mathbf{q},0}$ falling at normal incidence on the bilayer
graphene system with polarisation vector $\widehat{\mathbf{e}}_{\mathbf{p}}$%
, the absorption is:%
\begin{eqnarray}
P\left( \omega \right)  &=&\frac{E_{e}^{2}}{2}\Re\left[ \widehat{%
\mathbf{e}}_{p}^{\ast }\cdot \overleftrightarrow{\sigma }\left( 0,0,\omega
\right) \cdot \widehat{\mathbf{e}}_{p}\right]   \label{pomega} \\
&&-\frac{E_{e}^{2}}{2}\sum_{\mathbf{G}}\Re\left[ \frac{4\pi i}{\omega }%
\frac{1}{L}\sum_{q_{z}}\frac{\omega ^{2}}{\omega
^{2}-c^{2}q_{z}^{2}-c^{2}G^{2}}\right] \left\vert \overleftrightarrow{\sigma 
}\left( \mathbf{G},0,\omega \right) \cdot \widehat{\mathbf{e}}%
_{p}\right\vert ^{2}  \notag \\
&&+\frac{E_{e}^{2}}{2}\sum_{\mathbf{G}}\Re\left[ \frac{4\pi i}{\omega }%
\frac{1}{L}\sum_{q_{z}}\frac{c^{2}G^{2}}{\omega
^{2}-c^{2}q_{z}^{2}-c^{2}G^{2}}\right] \left\vert \widehat{\mathbf{G}}\cdot 
\overleftrightarrow{\sigma }\left( \mathbf{G},0,\omega \right) \cdot 
\widehat{\mathbf{e}}_{p}\right\vert ^{2}.  \notag
\end{eqnarray}%
The lattice constant in the crystal phase is $a_{0}=220$ \AA\ at $B=10$ T
(using the relation $2\pi n\ell ^{2}=\widetilde{\nu }=1$). It follows that
the smallest reciprocal lattice vector $G\approx 10^{8}$ m$^{-1}$ and the
corresponding frequency $cG\approx 10^{16}$ rad/s. The frequency of the
collective modes, on the other hand, is of order $10$ meV i.e. $\approx
10^{13}$ rad/s. Thus, $\omega <<cG.$ Keeping terms to order one in $\omega
/cG,$ we get for the absorption 
\begin{eqnarray}
P\left( \omega \right)  &\approx &\frac{E_{e}^{2}}{2}\Re\left[ 
\widehat{\mathbf{e}}_{p}^{\ast }\cdot \overleftrightarrow{\sigma }\left(
0,0,\omega \right) \cdot \widehat{\mathbf{e}}_{p}\right]   \label{tout} \\
&&+\frac{E_{e}^{2}}{2}\sum_{\mathbf{G}\neq 0}\Re\left[ \frac{2\pi i}{c}%
\left( \frac{\omega }{cG}\right) \right] \left\vert \overleftrightarrow{%
\sigma }\left( \mathbf{G},0,\omega \right) \cdot \widehat{\mathbf{e}}%
_{p}\right\vert ^{2}-\frac{E_{e}^{2}\pi }{c}\left\vert \overleftrightarrow{%
\sigma }\left( 0,0,\omega \right) \cdot \widehat{\mathbf{e}}_{p}\right\vert
^{2}  \notag \\
&&-\frac{E_{e}^{2}}{2}\sum_{\mathbf{G}}\Re\left[ \frac{2\pi i}{\omega }%
\right] \left\vert \mathbf{G}\cdot \overleftrightarrow{\sigma }\left( 
\mathbf{G},0,\omega \right) \cdot \widehat{\mathbf{e}}_{p}\right\vert ^{2}. 
\notag
\end{eqnarray}%
The only contributions at $\omega \neq 0$ come from the first and third
terms on the right-hand side of Eq. (\ref{tout}). Defining 
\begin{equation}
\overleftrightarrow{\sigma }\left( 0,0,\omega \right) =\frac{e^{2}}{h}%
\overleftrightarrow{\sigma }\left( \omega \right) ,  \label{sigma}
\end{equation}
we have 
\begin{equation}
P\left( \omega \neq 0\right) \approx \frac{e^{2}E_{e}^{2}}{2h}\Re\left[
\widehat{\mathbf{e}}_{p}^{\ast }\cdot \overleftrightarrow{\sigma }\left(
\omega \right) \cdot \widehat{\mathbf{e}}_{p}\right] -\alpha \frac{%
e^{2}E_{e}^{2}}{2h}\left\vert \overleftrightarrow{\sigma }\left( \omega
\right) \cdot \widehat{\mathbf{e}}_{p}\right\vert ^{2},  \label{abso2}
\end{equation}%
where $\alpha =e^{2}/\hslash c=1/137$ is the fine-structure constant and $%
\overleftrightarrow{\sigma }\left( \omega \right) $ is now unitless. For
realistic values of $\delta $, the condition $\alpha \overleftrightarrow{%
\sigma }\left( \omega \right) <<1$ is satisfied and the second term in the
right-hand side of Eq. (\ref{abso2}) can be neglected. The approximations
that we have made to get Eq. (\ref{abso2}) are equivalent to neglecting
retardation effects [i.e. to taking the limit $c\rightarrow \infty $ in Eq. (%
\ref{pomega})]. In fact, since retardation effects are neglected in the GRPA
for the conductivity, it seems logical to neglect them also in the
calculation of the absorption. We have finally at finite frequency%
\begin{equation}
P\left( \omega \right) =\frac{e^{2}E_{e}^{2}}{2h}\Re\left[ \widehat{%
\mathbf{e}}_{\mathbf{p}}^{\ast }\cdot \overleftrightarrow{\sigma }\left(
\omega \right) \cdot \widehat{\mathbf{e}}_{\mathbf{p}}\right] .  \label{abso}
\end{equation}

\end{widetext}

We take the incoming wave to be circularly polarized so that $\widehat{%
\mathbf{e}}_{p}\rightarrow \widehat{\mathbf{e}}_{\pm }=\left( \widehat{%
\mathbf{x}}\pm i\widehat{\mathbf{y}}\right) /\sqrt{2}$ and the absorption
is, from Eq. (\ref{abso}),%
\begin{equation}
P_{\pm }\left( \omega \right) =\frac{e^{2}E_{e}^{2}}{2h}\Re\left[
\sigma _{\pm }\left( \omega \right) \right] ,
\end{equation}%
where the conductivities $\sigma _{\pm }$ are defined by 
\begin{equation}
\sigma _{\pm }\equiv \sigma _{x,x}+\sigma _{y,y}\pm i\sigma _{x,y}\mp
i\sigma _{y,x}.
\end{equation}

Now, the optical conductivity is related to the current response $\Xi
_{\alpha ,\beta }\left( \tau \right) \equiv -\frac{1}{\hslash S}\left\langle
TJ_{\alpha }\left( \tau \right) J_{\beta }\left( 0\right) \right\rangle $ by%
\begin{eqnarray}
\Re\left[ \sigma _{i,i}\left( \omega \right) \right] &=&-\frac{\Im%
\left[ \Xi _{i,i}\left( \omega \right) \right] }{\omega },  \label{c1} \\
\Im\left[ \sigma _{i,i}\left( \omega \right) \right] &=&\frac{\Re%
\left[ \Xi _{i,i}\left( \omega \right) -\Xi _{i,i}\left( 0\right) \right] }{%
\omega },  \label{c2}
\end{eqnarray}%
and, for $i\neq j$, by 
\begin{equation}
\sigma _{i,j}\left( \omega \right) =\frac{i}{\left( \omega +i\delta \right) }%
\Xi _{i,j}\left( \omega \right)  \label{c3}
\end{equation}%
so that, for $\omega \neq 0$ and $i\neq j$, we have%
\begin{eqnarray}
\Re\left[ \sigma _{i,j}\left( \omega \right) \right] &=&-\frac{\Im%
\left[ \Xi _{i,j}\left( \omega \right) \right] }{\omega }, \\
\Im\left[ \sigma _{i,j}\left( \omega \right) \right] &=&\frac{\Re%
\left[ \Xi _{i,j}\left( \omega \right) \right] }{\omega }.
\end{eqnarray}

The total current operator is defined by $\mathbf{J}=-c\left. \partial
H^{0}/\partial \mathbf{A}^{e}\right\vert _{\mathbf{A}^{e}=0},$ where $H^{0}$
is the non-interacting Hamiltonian of Eq. (\ref{nonh}) where a Peierls
substitution has been made to take into account an external electromagnetic
field. In the two-component model, this current operator is related to the
pseudospin operator by

\begin{equation}
J_{\alpha }\left( \tau \right) =-\frac{\sqrt{2}e\ell N_{\varphi }}{\hslash }%
\left( \zeta -\beta \Delta _{B}\right) p_{\overline{\alpha }}\left( \tau
\right) ,
\end{equation}%
with the convention that $\overline{\alpha }=x$ if $\alpha =y$ and \textit{%
vice versa} and with $p_{\alpha }\left( \tau \right) =p_{\alpha }\left( 
\mathbf{q}=0,\tau \right) .$ The two-particle current Matsubara Green's
function tensor $\Xi _{\alpha ,\beta }$ evaluated at $\mathbf{q}=0$ can be
related to the two-particle pseudospin Matsubara Green's function tensor $%
K_{\alpha ,\beta }\left( \tau \right) =-\frac{N_{\varphi }}{\hslash }%
\left\langle Tp_{\alpha }\left( \tau \right) p_{\beta }\left( 0\right)
\right\rangle $ by

\begin{eqnarray}
\Xi _{\alpha ,\beta }\left( \tau \right) &\equiv &-\frac{1}{\hslash S}%
\left\langle TJ_{\alpha }\left( \tau \right) J_{\beta }\left( 0\right)
\right\rangle \\
&=&\frac{e^{2}}{\pi \hslash ^{2}}\left( \zeta -\beta \Delta _{B}\right) ^{2}%
\left[ -\frac{N_{\varphi }}{\hslash }\left\langle Tp_{\overline{\alpha }%
}\left( \tau \right) p_{\overline{\beta }}\left( 0\right) \right\rangle %
\right]  \notag \\
&\equiv &\frac{e^{2}}{\pi \hslash ^{2}}\left( \zeta -\beta \Delta
_{B}\right) ^{2}K_{\overline{\alpha },\overline{\beta }}\left( \tau \right) .
\notag
\end{eqnarray}%
Finally, $K_{\alpha ,\beta }\left( \tau \right) $ can easily be related to
the two-particle Green's functions defined with the operators in Eq. (\ref%
{order2}) i.e.

\begin{eqnarray}
\chi _{n_{1}n_{2}n_{3}n_{4}}\left( \tau \right) &=&-\frac{N_{\varphi }}{%
\hslash }\left\langle T\rho _{n_{1},n_{2}}\left( \tau \right) \rho
_{n_{3},n_{4}}\left( 0\right) \right\rangle \\
&&+\frac{N_{\varphi }}{\hslash }\left\langle \rho
_{n_{1},n_{2}}\right\rangle \left\langle \rho _{n_{3},n_{4}}\right\rangle 
\notag
\end{eqnarray}%
by the equations 
\begin{eqnarray}
K_{x,x} &=&\frac{1}{4}\left[ \chi _{++}+\chi _{+-}+\chi _{-+}+\chi _{--}%
\right] , \\
K_{y,y} &=&-\frac{1}{4}\left[ \chi _{++}-\chi _{+-}-\chi _{-+}+\chi _{--}%
\right] , \\
K_{x,y} &=&\frac{1}{4i}\left[ \chi _{++}-\chi _{+-}+\chi _{-+}-\chi _{--}%
\right] , \\
K_{y,x} &=&\frac{1}{4i}\left[ \chi _{++}+\chi _{+-}-\chi _{-+}-\chi _{--}%
\right] ,
\end{eqnarray}%
where, to shorten the notation, we have defined the associations $+\equiv 01$
and $-\equiv 10.$

The current response is not isotropic in all phases so that $K_{x,x}\neq K_{y,y}$ in general. 
For the conductivities at $%
\mathbf{q}=0,$%
\begin{eqnarray}
\Re\left[ \sigma _{+}\left( \omega \right) \right] &=&-\frac{e^{2}}{h}%
\frac{\left( \zeta -\beta \Delta _{B}\right) ^{2}}{\hslash \omega }\Im%
\left[ \chi _{+-}\left( \omega \right) \right] , \\
\Re\left[ \sigma _{-}\left( \omega \right) \right] &=&-\frac{e^{2}}{h}%
\frac{\left( \zeta -\beta \Delta _{B}\right) ^{2}}{\hslash \omega }\Im%
\left[ \chi _{-+}\left( \omega \right) \right] ,
\end{eqnarray}%
and for the absorption per unit area 
\begin{eqnarray}
P_{+}\left( \omega \right) &=&-\frac{e^{2}E_{e}^{2}}{2h}\frac{\left( \zeta
-\beta \Delta _{B}\right) ^{2}}{\hslash \omega }\Im\left[ \chi
_{+-}\left( \omega \right) \right] ,  \label{absop} \\
P_{-}\left( \omega \right) &=&-\frac{e^{2}E_{e}^{2}}{2h}\frac{\left( \zeta
-\beta \Delta _{B}\right) ^{2}}{\hslash \omega }\Im\left[ \chi
_{-+}\left( \omega \right) \right] .  \label{absom}
\end{eqnarray}

For linear polarizations, the absorption is given by

\begin{eqnarray}
P_{x}\left( \omega \right) &=&\frac{e^{2}E_{e}^{2}}{2h}\frac{\left( \zeta
-\beta \Delta _{B}\right) ^{2}}{\hslash \omega } \\
&&\times \Im\left[ \chi _{++}\left( \omega \right) -\chi _{+-}\left(
\omega \right) -\chi _{-+}\left( \omega \right) +\chi _{--}\left( \omega
\right) \right] ,  \notag \\
P_{y}\left( \omega \right) &=&-\frac{e^{2}E_{e}^{2}}{2h}\frac{\left( \zeta
-\beta \Delta _{B}\right) ^{2}}{\hslash \omega } \\
&&\times \Im\left[ \chi _{++}\left( \omega \right) +\chi _{+-}\left(
\omega \right) +\chi _{-+}\left( \omega \right) +\chi _{--}\left( \omega
\right) \right] ,  \notag
\end{eqnarray}%
for an electric field polarized in the $\widehat{\mathbf{x}}$ or $\widehat{%
\mathbf{y}}$ direction respectively.

\subsection{Incoherent phases $I$ and $I^{\ast }$}

Analytical expressions for the absorption are possible in the incoherent
phases when the parallel electric field $\mathbf{E}_{\Vert }=0$. Using the
results derived in Ref. \onlinecite{ReneSU8}, the only two non-zero GRPA\
response functions at $\mathbf{q}=0$ in these phases are $\chi _{+-}\left(
\omega \right) $ and $\chi _{-+}\left( \omega \right) $ with%
\begin{equation}
\chi _{\mp \pm }\left( \omega \right) =\pm \frac{\nu _{1}-\nu _{0}}{\hslash
\omega +i\delta \pm \left( \zeta -\beta \Delta _{B}+\frac{1}{4}\eta \nu
_{1}\right) }.  \label{chi}
\end{equation}%
For phases $I$ and $I^{\ast }$, the absorption is given by%
\begin{eqnarray}
P_{\pm }^{\left( I\right) }\left( \omega \right)  &=&\frac{e^{2}E_{e}^{2}}{2h%
}\frac{\pi }{\hslash }\beta \left( \Delta _{O}-\Delta _{B}\right) \delta
\left( \omega \mp \omega _{I}\left( 0\right) \right) ,  \label{inco1} \\
P_{\pm }^{\left( I^{\ast }\right) }\left( \omega \right)  &=&\frac{%
e^{2}E_{e}^{2}}{2h}\frac{\pi }{\hslash }\beta \frac{\left( \Delta
_{B}-\Delta _{O}\right) ^{2}}{\left( \Delta _{B}-\Delta _{O^{\ast }}\right) }%
\delta \left( \omega \pm \omega _{I^{\ast }}\left( 0\right) \right) ,
\label{inco3}
\end{eqnarray}%
where the frequencies $\omega _{I}\left( 0\right) $ (plotted in Fig. 8) and $%
\omega _{I^{\ast }}\left( 0\right) $ have been defined in Sec. III. We
remark that, from Eq. (\ref{chi}), it is clear that there is no contribution
from the bubble (or polarization) diagrams in $\chi _{\mp \pm }\left( \omega
\right) $ so that $\overleftrightarrow{\widetilde{\sigma }}\left( 0,0,\omega
\right) =\overleftrightarrow{\sigma }\left( 0,0,\omega \right) $ in these
phases. This is not true in the nonuniform phases however.

The pseudospin-wave mode in these phases is circularly polarized. It shows
up \textit{only} in $P_{+}\left( \omega \right) $ for the $I$ phase and 
\textit{only }in $P_{-}\left( \omega \right) $ for the $I^{\ast }$ phase.
The response functions $\chi _{+-}\left( \omega \right) $ and $\chi
_{-+}\left( \omega \right) $ are equal for conjugate phases, but because of
the prefactor in Eqs. (\ref{absop})-(\ref{absom}), the intensity of the
absorption is not. The absorption $P_{+}\left( \omega \right) $ decreases as 
$\Delta _{B}\rightarrow $ $\Delta _{O}$ in the $I$ phase while $P_{-}\left(
\omega \right) $ diverges (in the absence of disorder) at the $O^{\ast
}\rightarrow I^{\ast }$ transition, decreases as the bias is increased and
then increases linearly with $\Delta _{B}$ at still larger bias.

The maximal value of the absorption frequency $\omega _{I}\left( 0\right) $
is at $\Delta _{B}=0$ where $\hslash \omega _{I}\left( 0\right) =3.85$ meV
i.e. $\nu _{I}\left( 0\right) =9.3\times 10^{11}$ Hz in the far infrared.
This frequency can be tuned all the way to zero by increasing $\Delta _{B}.$
It can also be increased by a finite $\mathbf{E}_{\Vert }$ as shown in Fig.
8. With a finite $\mathbf{E}_{\Vert },$ the absorption still appears
predominantly in $P_{+}\left( \omega \right) $ for phase $I$ and in $%
P_{-}\left( \omega \right) $ for phase $I^{\ast },$ but the other circular
component makes a very small contribution in each case.

\subsection{Coherent phases $O$ and $O^{\ast }$}

In the coherent phases $O$ and $O^{\ast },$ the pseudospin mode is gapless
and there is no absorption at finite frequency. As shown in Fig. 6, a finite 
$\mathbf{E}_{\Vert }$ gaps that mode and makes it visible in absorption. The
range in bias where these two phases are the ground state is so small that
the optical frequency can't change much with bias. The absorption is
predominantly in $P_{+}\left( \omega \right) $ for $\Delta _{B}<\Delta _{M}$
and in $P_{-}\left( \omega \right) $ for $\Delta _{B}>\Delta _{M}$ but there
is a small intensity in the other circular polarization which is smaller by
a factor $\approx 10.$

\subsection{Crystal phases $C$ and $C^{\ast }$}

Figure 10 shows the absorptions $P_{\pm }\left( \omega \right) $ in the
conjugate crystal phases $C$ [Fig. 10(a),(b))]\ and $C^{\ast }$ [Fig.
10(c),(d)] at $\Delta _{B}=66$ meV and $\Delta _{B}^{\ast }=193.8$ meV with $%
E_{\Vert }=0$ and $E_{\Vert }=0.04$ mV/nm. This figure should be compared
with Fig. 4 (c) where the dispersion of the collective modes of the crystal
is plotted. The gapless phonon mode is absent of the spectrum for $\omega
>0. $ As in the $I$ and $I^{\ast }$ phase, the modes active in $P_{+}\left(
\omega \right) $ for $\Delta _{B}<\Delta _{M}$ are those active in $%
P_{-}\left( \omega \right) $ for $\Delta _{B}>\Delta _{M}$ and vice versa.
The absorption peaks are more intense when $\Delta _{B}>\Delta _{M}$
however. For $\Delta _{B}<\Delta _{M},$ the first gapped mode is seen only
in $P_{+}\left( \omega \right) $ while the fourth one is seen only in $%
P_{-}\left( \omega \right) $ and is more than ten times smaller in
intensity. The second and third modes are not active in $P_{\pm }\left(
\omega \right) $ or $P_{x,y}\left( \omega \right) $ (not shown in the
figure)\ for $E_{\Vert }=0.$ A finite $E_{\Vert }$ activates the second and
third modes but they get a very small intensity: the second mode is then
active in $P_{-}\left( \omega \right) $ only while the third mode, showing
up in both polarizations, seems to be linearly polarized. The energy of the
dominant mode is $\hslash \omega \approx 4$ meV i.e. similar to the energy
of the pseudospin mode at zero bias in the $I$ phase. A parallel electric
field does not change noticeably the frequency as we noted before. A
linearly polarized electromagnetic wave excites all modes of $P_{+}\left(
\omega \right) $ and $P_{-}\left( \omega \right) $. In the crystal phases,
we have checked that the absorption $P_{x}\left( \omega \right) =P_{y}\left(
\omega \right) .$

%%%%%%%%%%%%%%%%%%%%%%%%%%%%%%%%%%%%%%%%%%%%%%%%%%%%%%%%%%%%%%%%%%%%%%%%%%
\begin{figure}[tbph]
\includegraphics[scale=1.0]{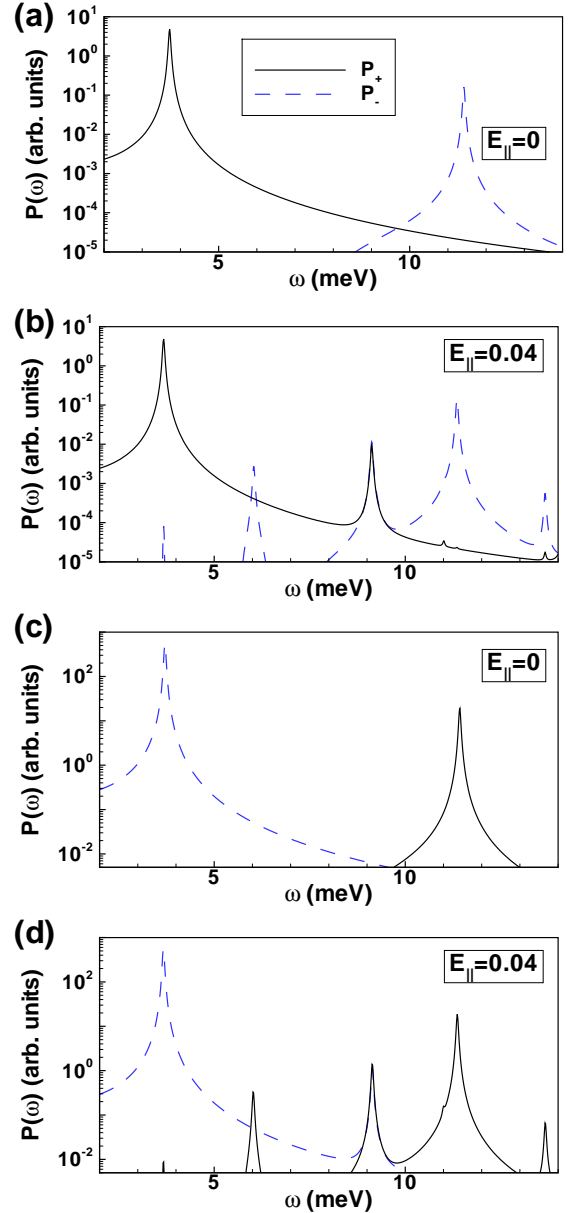}
\caption{(Color online) Absorptions $%
P_{\pm }\left( \protect\omega \right) $ in the crystal phases $C$ at $\Delta
_{B}=66$ meV [(a) and (b)] and $C^{\ast }$ at the conjugate bias $\Delta
_{B}=194$ meV [(c) and (d)] with and without a parallel electric field $%
E_{\Vert }=0.04$ mV/nm.}
\label{figure10}
\end{figure}
%%%%%%%%%%%%%%%%%%%%%%%%%%%%%%%%%%%%%%%%%%%%%%%%%%%%%%%%%%%%%%%%%%%%%%%%%%

\subsection{Spiral phase}

Figure 1 shows the absorptions $P_{\pm }\left( \omega \right) $ in the
spiral phase at conjugate biases $\Delta _{B}=77$ meV [Fig. 11(a),(b)] and $%
\Delta _{B}^{\ast }=183$ meV [Fig. 11(c),(d)] with and without and electric
field $E_{\Vert }=0.04$ mV/nm. This figure should be compared with Fig. 4
(d) where the collective modes of the spiral are plotted. The gapless phonon
mode is absent of the spectrum for $\omega >0.$ The spectrum has the same
symmetry in $P_{\pm }\left( \omega \right) $ as in the crystal phases. For $%
\Delta _{B}<\Delta _{M},$ the first gapped mode is seen predominantly in $%
P_{+}\left( \omega \right) $ while the almost degenerate third and fourth
gapped mode are more active in $P_{-}\left( \omega \right) .$ In contrast
with the crystal phase, however, the modes of the spiral are not fully
circularly polarized since the first(third/fourth) mode is only about $5$
times stronger(weaker) in $P_{+}\left( \omega \right) $ than in $P_{-}\left(
\omega \right) .$ In linear polarization (not shown in Fig. 4), both modes
are stronger in $P_{y}\left( \omega \right) $ that in $P_{x}\left( y\right)
. $ Note that the spiral rotates in the $z-x$ plane for $E_{\Vert }=0$ and
the absorption is maximal when the electric field of the electromagnetic
field is perpendicular to the spiral i.e. in $P_{y}\left( \omega \right) $,
a fact already mentioned in Ref. \onlinecite{ReneHelical}. The energy of the
most intense mode is $\approx 9$ meV, slightly higher than in the crystal.
At $\Delta _{B}=\Delta _{M},$ the two most active modes are fully linearly
polarized: the first in $y$ and the second in $x$ as shown in Fig. 11(e).
Thus, these two modes become more linearly polarized as $\Delta
_{B}\rightarrow \Delta _{M}$ from above or from below.

The second mode is not active in absorption when $E_{\Vert }=0$ but is
activated when $E_{\Vert }\neq 0$. With a finite $E_{\Vert },$ the
degeneracy of the third and fourth modes is lifted and they become active in
absorption. The second mode is fully linearly polarized and appears in $%
P_{y}\left( \omega \right) $.

%%%%%%%%%%%%%%%%%%%%%%%%%%%%%%%%%%%%%%%%%%%%%%%%%%%%%%%%%%%%%%%%%%%%%%%%%%
\begin{figure}[tbph]
\includegraphics[scale=1.08]{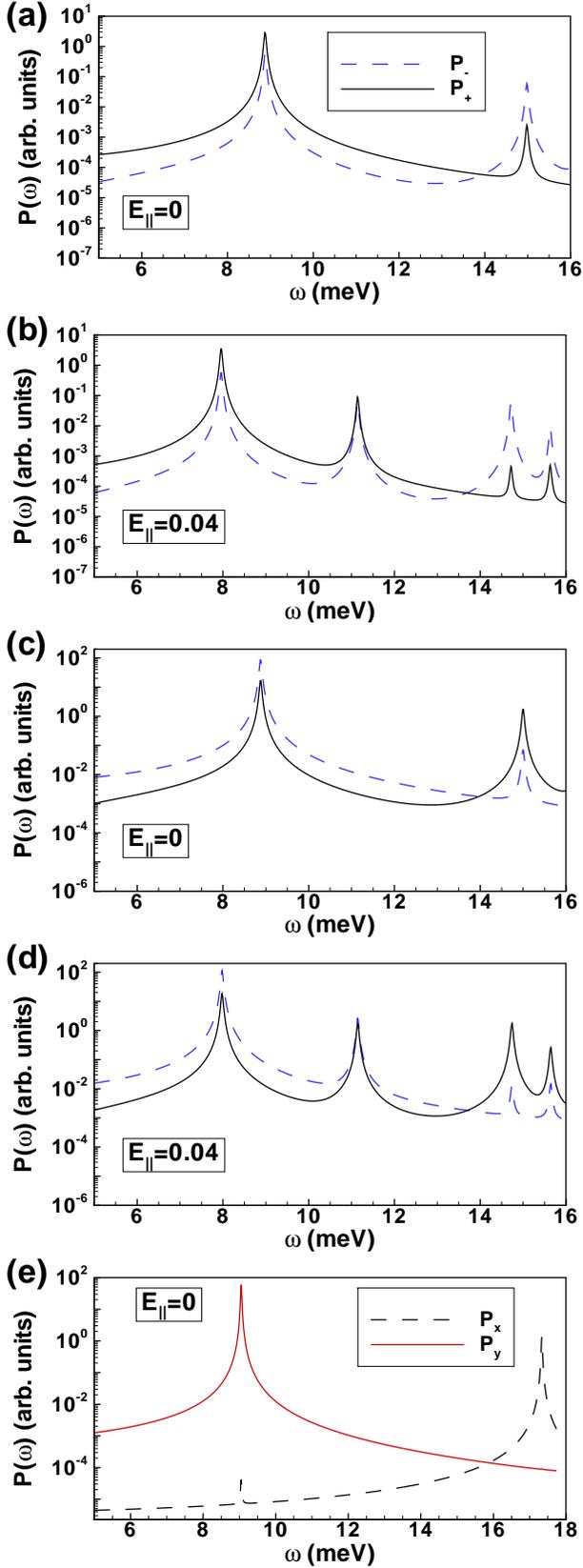}
\caption{(Color online) Absorptions $%
P_{\pm }\left( \protect\omega \right) $ in the spiral phases at conjugate
biases $\Delta _{B}=77$ meV [(a) and (b)] and $\Delta _{B}=183$ meV [(c) and
(d)] with and without a parallel electric field $E_{\Vert }=0.04$ mV/nm. (e)
Absorption at bias $\Delta _{M}=130$ meV and for $E_{\Vert }=0.$ }
\label{figure11}
\end{figure}
%%%%%%%%%%%%%%%%%%%%%%%%%%%%%%%%%%%%%%%%%%%%%%%%%%%%%%%%%%%%%%%%%%%%%%%%%%

\section{KERR\ ROTATION FROM COLLECTIVE MODES}

We adapt the calculation of Ref. \onlinecite{CalculKerr} (see Supplemental
material of this reference) to our specific problem to compute the Kerr
angle for a graphene bilayer on top of a dielectric with a refractive index $%
n_{2}$. We do not assume that $\sigma _{xx}\left( \omega \right) =\sigma
_{yy}\left( \omega \right) $ since this symmetry is not satisfied in all
phases and do not make the simplifying assumptions $\frac{4\pi }{c}\sigma
_{xx}\left( \omega \right) <<1,\frac{4\pi }{c}\sigma _{xy}\left( \omega
\right) <<1$. We take the incident wave $\mathbf{E}_{I}$ in the medium $1$
(the vacuum) to be linearly polarized. We write for the incident, reflected
and transmitted (in medium $2$) waves 
\begin{eqnarray}
\mathbf{E}_{I} &=&E_{I}\widehat{\mathbf{x}}e^{i\left( k_{1}z-\omega t\right)
}, \\
\mathbf{E}_{R} &=&E_{I}\left( r_{xx}\widehat{\mathbf{x}}+r_{yx}\widehat{%
\mathbf{y}}\right) e^{i\left( -k_{1}z-\omega t\right) }, \\
\mathbf{E}_{T} &=&E_{I}\left( t_{xx}\widehat{\mathbf{x}}+t_{yx}\widehat{%
\mathbf{y}}\right) e^{i\left( k_{2}z-\omega t\right) },
\end{eqnarray}%
and for the corresponding magnetic fields%
\begin{eqnarray}
\mathbf{B}_{I} &=&\widehat{\mathbf{z}}\times \mathbf{E}_{I}=E_{I}\widehat{%
\mathbf{y}}e^{i\left( k_{1}z-\omega t\right) }, \\
\mathbf{B}_{R} &=&-\widehat{\mathbf{z}}\times \mathbf{E}_{R}=E_{I}\left(
-r_{xx}\widehat{\mathbf{y}}+r_{yx}\widehat{\mathbf{x}}\right) e^{i\left(
-k_{1}z-\omega t\right) }, \\
\mathbf{B}_{T} &=&n_{2}\widehat{\mathbf{z}}\times \mathbf{E}%
_{T}=n_{2}E_{I}\left( t_{xx}\widehat{\mathbf{y}}-t_{yx}\widehat{\mathbf{x}}%
\right) e^{i\left( k_{2}z-\omega t\right) },
\end{eqnarray}%
with $k_{1}=\omega /c$ and $k_{2}=\omega n_{2}/c.$ The wave arrives at
normal incidence on the bilayer graphene and is in part reflected in medium
1 and in part transmitted to medium $2.$

The boundary condition~$\mathbf{E}_{2}^{\Vert }=\mathbf{E}_{1}^{\Vert }$
gives the equation%
\begin{equation}
E_{I}\widehat{\mathbf{x}}+E_{I}\left( r_{xx}\widehat{\mathbf{x}}+r_{yx}%
\widehat{\mathbf{y}}\right) =E_{I}\left( t_{xx}\widehat{\mathbf{x}}+t_{yx}%
\widehat{\mathbf{y}}\right) ,  \label{sol2}
\end{equation}%
while the boundary condition (the system is non-magnetic so that we take $%
\mathbf{B}=\mathbf{H}$) $\mathbf{B}_{2}^{\Vert }-\mathbf{B}_{1}^{\Vert }=%
\frac{4\pi }{c}\mathbf{j}\times \widehat{\mathbf{z}}$ gives%
\begin{eqnarray}
&&\mathbf{\ }n_{2}E_{I}\left( t_{xx}\widehat{\mathbf{y}}-t_{yx}\widehat{%
\mathbf{x}}\right) -E_{I}\widehat{\mathbf{y}}+E_{I}\left( r_{xx}\widehat{%
\mathbf{y}}-r_{yx}\widehat{\mathbf{x}}\right)  \label{sol1} \\
&=&\frac{4\pi }{c}\left( j_{y}\widehat{\mathbf{x}}-j_{x}\widehat{\mathbf{y}}%
\right) ,  \notag
\end{eqnarray}%
where $\mathbf{j}$ is the induced surface current in the graphene bilayer.
Now, according to our definition of the screened conductivity in Eq. (\ref%
{cond2}), we must take 
\begin{equation}
\mathbf{j}\left( \omega \right) =\frac{e^{2}}{h}\overleftrightarrow{\sigma }%
\left( \omega \right) \cdot \mathbf{E}_{I}\left( z=0,\omega \right) ,
\end{equation}%
where $\overleftrightarrow{\sigma }\left( \omega \right) $ is defined in Eq.
(\ref{sigma}). We implicitly assumed in the above derivation that, in the
crystal and spiral phases, we can neglect the reent components $\left\{ 
\mathbf{j}\left( \mathbf{G},\omega \right) \right\} $ and keep only $\mathbf{%
j}\left( \omega \right) \equiv $ $\mathbf{j}\left( \mathbf{G=0},\omega
\right) $ and neglect the diffracted components in the electric field as
well.

Solving Eqs. (\ref{sol1},\ref{sol2}), we find for the Kerr
(counter-clockwise) rotation angle from the $x$ axis:

\begin{eqnarray}
\tan \left( \theta _{K}\right)  &=&\Re\left( \frac{r_{yx}}{r_{xx}}%
\right)   \label{kerr} \\
&=&2\alpha \Re \left[ \frac{\sigma _{yx}\left( \omega \right) }%
{n_{2}-1+2\alpha \sigma _{xx}\left( \omega \right) }\right]   \notag
\end{eqnarray}%
where $\alpha $ is the fine-structure constant. For an incident wave
polarized along the $y$ axis, the (clockwise) rotation angle is obtained by
the substitution $\sigma _{xx}\rightarrow \sigma _{yy}$ in the denominator
of Eq. (\ref{kerr}).

Figure 12 shows the Kerr angle $\theta _{K}\left( \omega \right) $ for $%
\Delta _{B}=0$ in the $I$ phase and for its conjugate bias $\Delta _{B}=260$
meV in the $I^{\ast }$ phase. We take $n_{2}=\sqrt{5}$ and $E_{\Vert }=0.$
The Kerr effect occurs at the frequency of the collective mode which is the
same for both biases but the sense of rotation is opposite in the two
phases. The rotation in the $I^{\ast }$ phase takes place in a larger domain
of frequencies (i.e. it is slower) than in the $I$ phase and the maximum
Kerr angle is also bigger$.$ This is due to the prefactor $\left( \zeta
-\beta \Delta _{B}\right) ^{2}$ in the equation for the conductivity which
makes the conductivities much larger in phase $I^{\ast }$ and increases the
contribution of $\sigma _{xx}\left( \omega \right) $ in the denominator of
Eq. (\ref{kerr}).

The maximal Kerr angle is large in Fig. 12 because we have taken a very
small value for $\delta $ i.e. $\delta =0.014$ meV. For $\delta =0.14$ meV,
the maximum angle for $\Delta _{B}=0$ is reduced to $4.3$ degrees which is
of the order of the Kerr angle found for other types of broken-symmetry
states studied before.\cite{GorbarFaraday} For $\delta =1.4$ meV, it is
reduced to $0.7$ degrees. Since $\delta $ approximates the effect of
disorder, we see that the maximum Kerr angle is very sensitive to this
parameter.

We remark that conductivity $\sigma _{xy}\left( \omega \right) $ should
satisfy the condition $\Re\left[ \sigma _{xy}\left( \omega =0\right) %
\right] =-1$ in the two-level system since $\widetilde{\nu }=1$. This
condition is actually satisfied in phase $I$ only and not in all the other
phases. Indeed, Eq. (\ref{chi}) gives 
\begin{eqnarray}
\lim_{\delta \rightarrow 0}\sigma _{xy}^{\left( I\right) }\left( 0\right)
&=&-1, \\
\lim_{\delta \rightarrow 0}\sigma _{xy}^{\left( I^{\ast }\right) }\left(
0\right) &=&\frac{\left( \Delta _{B}-\Delta _{O}\right) ^{2}}{\left( \Delta
_{B}-\Delta _{O^{\ast }}\right) ^{2}}.
\end{eqnarray}%
We have verified that, when levels $\left\vert N\right\vert >0$ are included
in the calculation (in the four-component model and in the absence of
interaction), the condition $\Re\left[ \sigma _{xy}\left( \omega
=0\right) \right] =-3$ is then satisfied ($\widetilde{\nu }=1\Rightarrow $ $%
\nu =3$). The transitions from the levels $N\neq 0$ occur at higher
frequencies, however, and they should not affect much the behavior of the
Kerr rotation near a collective mode resonance a finite frequency.

%%%%%%%%%%%%%%%%%%%%%%%%%%%%%%%%%%%%%%%%%%%%%%%%%%%%%%%%%%%%%%%%%%%%%%%%%%
\begin{figure}[tbph]
\includegraphics[scale=1.0]{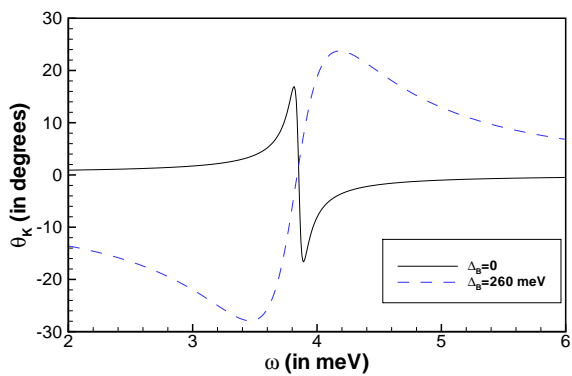}
\caption{(Color online) Kerr angle for
conjugate biases $\Delta _{B}=0$ and $\Delta _{B}=260$ meV in the \ $I$ and $%
I^{\ast }$ phases.}
\label{figure12}
\end{figure}
%%%%%%%%%%%%%%%%%%%%%%%%%%%%%%%%%%%%%%%%%%%%%%%%%%%%%%%%%%%%%%%%%%%%%%%%%%

Figure 13 shows the Kerr angle in the crystal and spiral phases for
conjugate biases: (a) $\Delta _{B}=66$ meV and (b) $\Delta _{B}=194$ meV in
the crystal phase and (c) $\Delta _{B}=77$ meV and (d) $\Delta _{B}=183$ meV
in the spiral phase. We take $n_{2}=\sqrt{5}$ and $E_{\Vert }=0.$ As in
phase $I$ and $I^{\ast }$ discussed above, our calculation does not include
disorder so that we cannot get a quantitative result for the Kerr angle. We
use $\delta \approx 0.014$ meV in all curves in this figure and use the same
step in frequency so that we can compare the relative size of the rotation
angle. As in Fig. 12, we expect that if disorder is included, it will reduce
the Kerr angle. The modes visible in the Kerr rotation are also those active
in absorption as can be seen by comparing Fig. 13 with Figs. 10-11. In the
crystal and spiral phases where two modes are active in the range of
frequency shown in Fig. 12, the Kerr angle for $\Delta _{B}<\Delta _{M}$
increases in going towards the resonance in the lower-energy mode while it
decreases when approaching the fourth mode (the second and third mode do not
appear in the Kerr rotation). The rotations are in the opposite directions
for the conjugate biases. The Kerr rotation is bigger by a factor $\approx 10
$ in the conjugate phases with $\Delta _{B}>\Delta _{M}.$ Since the same
mode leads to opposite rotation of the polarization in for conjugate biases,
the rotation must cease at bias $\Delta _{^{M}}$. We have checked
numerically that this is the case.

In all phases but the spiral phase, we get numerically that the Kerr angle
is the same for an incident wave polarized along the $x$ or $y$ axis. This
is also true at small bias ($\Delta _{B}<\Delta _{M}$) in the spiral phase:
a $y$ polarization gives the same result as in Fig. 13 (c).  For $\Delta
_{B}>\Delta _{M},$ however, the rotation angle is bigger when the
polarization is along the $x$ axis (i.e. for a spiral rotating in the $z-x$
plane) as can be seen by comparing Fig. 13 (d) and Fig. 13 (e). 

%%%%%%%%%%%%%%%%%%%%%%%%%%%%%%%%%%%%%%%%%%%%%%%%%%%%%%%%%%%%%%%%%%%%%%%%%%
\begin{figure}[tbph]
\includegraphics[scale=1.0]{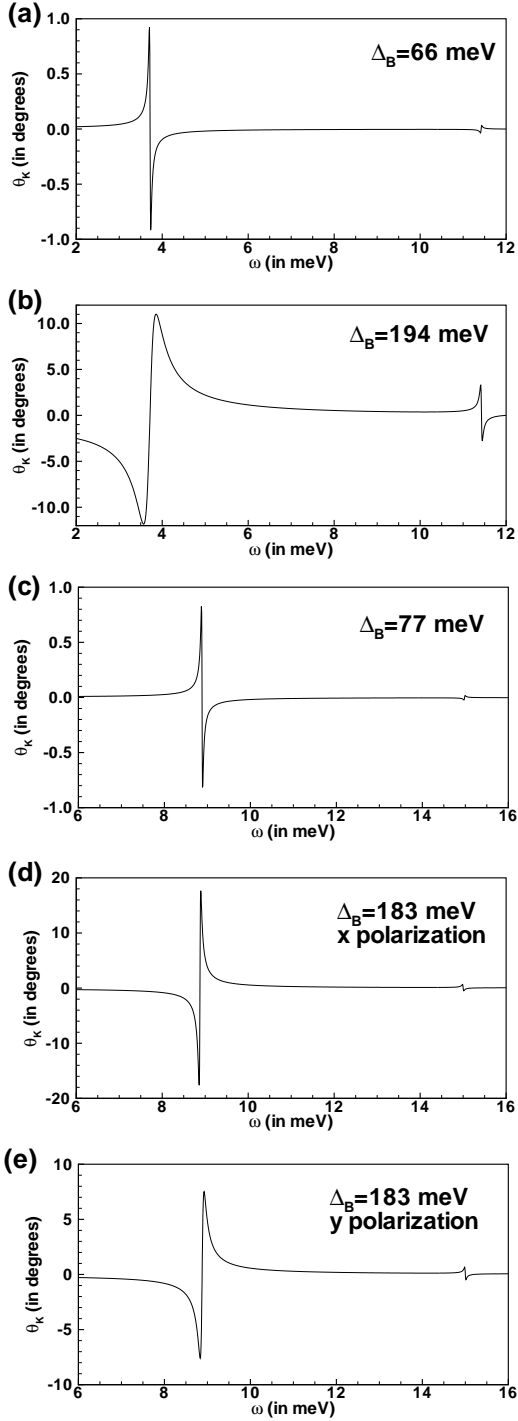}
\caption{Kerr angle for conjugate
biases (a) $\Delta _{B}=66$ meV and (b) $\Delta _{B}=194$ meV in the crystal
phase and conjugate biases (c) $77$ meV and (d) $183$ meV in the spiral
phase with a $x$ polarization and (e) with a $y$ polarization of the
incident wave.}
\label{figure13}
\end{figure}
%%%%%%%%%%%%%%%%%%%%%%%%%%%%%%%%%%%%%%%%%%%%%%%%%%%%%%%%%%%%%%%%%%%%%%%%%%

\section{CONCLUSION}

We have continued in this work an analysis of a sequence of phase
transitions involving uniform and nonuniform states with orbital coherence
initiated in Ref. \onlinecite{ReneHelical}. This sequence, represented in
Fig. 1, occurs at filling factor $\nu =3$ of bilayer graphene, in Landau
level $N=0,$ when an electric bias between the two levels is increased. At
filling factor $\nu =1,$ the same states are found in the phase diagram but
others involving spin coherence are present as well. In this work, we choose
to concentrate on the simpler case of $\nu =3$ in order to simplify the
discussion but the same results would apply to the corresponding states at $%
\nu =1.$

Because we were only interested in studying the signature of the collective
excitations due to transitions between the two levels $n=0,1$ in Landau
level $N=0,$ we could use the two-component model\cite{McCann} to simplify
the calculations instead of working with the full four band model of bilayer
graphene. All calculations were done at $T=0$ K using the Hartree-Fock
approximation to derive the phase diagram of the interacting chiral electron
gas and the generalized random-phase approximation (GRPA)\ to obtain the
dispersion relations of the collective modes.

Our calculations show that there are qualitative differences in the
transport gap and in the optical properties (electromagnetic absorption and
Kerr rotation) in the different phases represented in Fig. 1. The absorption
frequency can be tuned all the way to zero in the incoherent phase while it
is almost constant in the crystal and spiral phase and zero in the uniform
state with orbital coherence. The absorption frequency can be modified by
applying an electric field in the plane of the layers, but in a limited way
since such a field suppresses the nonuniform states. Another clear signature
occurs in what we call the conjugate states where the same optically active
modes at two conjugate biases are active in opposite circular polarizations.
Moreover, the optically active modes in the incoherent and crystal phases
are circularly polarized in contrast to the active modes in the spiral phase
which are neither completely linearly polarized nor completely circularly
polarized.

Another qualitative difference that we find is in the behavior of the Kerr
rotation. There is a Kerr effect near the frequency of each mode active in
the absorption. Conjugate states show a Kerr effect at the same frequency
but with opposite sign for the rotation of the polarization. The maximal
polarization angle (near the resonant frequency) is bigger for bias $\Delta
_{B}$ above the middle of the spiral phase. The observability of the Kerr
effect would depend very much on the strength of the disorder in the sample.
A proper treatment of disorder is also necessary to derive the actual
absorption line shape due to the collective excitations.

The sequence of phase transitions that we studied in this paper has not been
observed so far. We think that the work presented here may help in
distinguishing the different phases. Of course, several improvements would
be necessary to get numerically accurate predictions. Landau level mixing,
for example, is known to modify the dispersion and hence the absorption
frequency of the collective modes or magnetoexcitons\cite{Toke}. Also, an
important information as regards the experimental observability of the
phases would be to know how fragile they are with respect to thermal
fluctuations. We leave these problems for further work.

\begin{acknowledgments}
R. C\^{o}t\'{e} was supported by a grant from the Natural Sciences and
Engineering Research Council of Canada (NSERC). Computer time was provided
by Calcul Qu\'{e}bec and Compute Canada.
\end{acknowledgments}


\begin{thebibliography}{99}
\bibitem{BilayerReview} For a review of some of the properties of graphene
and bilayer graphene, see for example: A. H. Castro Neto, F. Guinea, N. M.
R. Peres, K. S. Novoselov and A. K. Geim, Rev. Mod. Phys. \textbf{81}, 109
(2009); D. S. L. Abergel, V. Apalkov, J. Berashevich, K. Ziegler and Tapash
Chakraborty, Advances in Physics \textbf{59}, 261 (2010); M. O. Goerbig,
Rev. Mod. Phys. \textbf{83}, 1193 (2011); Edward McCann and Mikito Koshino,
Rep. Prog. Phys. \textbf{76}, 056503 (2013).

\bibitem{BarlasReview} For a review of the C2DEG in bilayer graphene in
Landau level $N=0$, see for example: Yafis Barlas, Kun Yang, and A. H.
MacDonald, Nanotechnology \textbf{23}, 052001 (2012).

\bibitem{ReneSU8} J. Lambert and R. C\^{o}t\'{e}, Phys. Rev. B \textbf{87},
115415 (2013).

\bibitem{Gorbar} E. V. Gorbar, V. P. Gusynin, Junji Jia, and V. A. Miransky,
Phys. Rev. B \textbf{84}, 235449 (2011); E. V. Gorbar, V. P. Gusynin, and V.
A. Miransky, JETP Lett. \textbf{91}, 314 (2010); E. V. Gorbar, V. P.
Gusynin, and V. A. Miransky, Phys. Rev. B \textbf{81}, 155451 (2010); C. T%
\"{o}ke and V. I. Fal'ko, Phys. Rev. B \textbf{83}, 115455 (2011); E. V.
Gorbar, V. P. Gusynin, V. A. Miransky, and I. A. Shovkovy, Phys. Rev. B 
\textbf{85}, 235460 (2012).

\bibitem{ReneIsing} Wenchen Luo, R. C\^{o}t\'{e}, and Alexandre B\'{e}%
dard-Vall\'{e}e, Phys. Rev. B \textbf{90}, 075425 (2014).

\bibitem{BarlasPRL} Yavis Barlas, R. C\^{o}t\'{e}, K. Nomura, and A. H.
MacDonald, Phys. Rev. Lett. \textbf{101}, 097601 (2008).

\bibitem{Experiences} Benjamin E. Feldman, Jens Martin, and Amir Yacoby,
Nat. Phys. \textbf{6}, 889 (2009); Y. Zhao, P. Cadden-Zimansky, Z. Jiang,
and P. Kim, Phys. Rev. Lett. \textbf{104}, 066801 (2010); R. T. Weitz, M. T.
Allen, B. E. Feldman, J. Martin, and A. Yacoby, Science \textbf{330}, 812
(2010); J. Martin, B. E. Feldman, R. T. Weitz, M. T. Allen, and A. Yacoby,
Phys. Rev. Lett. \textbf{105}, 256806 (2010). Seyoung Kim, Kayoung Lee, and
E. Tutuc, Phys. Rev. Lett. \textbf{107}, 016803 (2011); Seyoung Kim, Kayoung
Lee, and E. Tutuc, Phys. Rev. Lett. \textbf{107}, 016803 (2011); J. Velasco
Jr, L. Jing, W. Bao, Y. Lee, P. Kratz, V. Aji, M. Bockrath, C. N. Lau, C.
Varma, R. Stillwell, D. Smirnov, Fan Zhang, J. Jung, and A. H. Macdonald,
Nat. Nanotechnol. \textbf{7}, 156 (2012); Seyoung Kim, Insun Jo, D. C.
Dillen, D. A. Ferrer, B. Fallahazad, Z. Yao, S. K. Banerjee, and E. Tutuc,
Phys. Rev. Lett. \textbf{108}, 116404 (2012); H. J. Elferen, A. Veligura, E.
V. Kurganova, U. Zeitler, J. C. Maan, N. Tombros, I. J. Vera-Marun, and B.
J. van Wees, Phys. Rev. B \textbf{85}, 115408 (2012). P. Maher, C. R. Dean,
A. F. Young, T. Taniguchi, K. Watanabe, K. L. Shepard, J. Hone and P. Kim,
Nature Phys. \textbf{9}, 154 (2013); Kayoung Lee, Babak Fallahazad, Hongki
Min, and Emanuel Tutuc, IEEE Trans. Electron Devices \textbf{60}, 103
(2013); K. Lee, B. Fallahazad, J. Xue, D. C. Dillen, K. Kim, T. Taniguchi,
K. Watanabe, and E. Tutuc, Science \textbf{345}, 58 (2014).

\bibitem{ReneOrbital} R. C\^{o}t\'{e}, Jules Lambert, Yafis Barlas, and A.
H. MacDonald, Phys. Rev. B \textbf{82}, 035445 (2010)

\bibitem{ReneHelical} R. C\^{o}t\'{e}, J. P. Fouquet, and Wenchen Luo, Phys.
Rev. B \textbf{84}, 235301 (2011).

\bibitem{HelicalMagnets} Jung Hoon Han, Jiadong Zang, Zhihua Yang, Jin-Hong
Park, and Naoto Nagaosa, Phy. Rev. B \textbf{82}, 094429 (2010); Jin-Hong
Park and Jung Hoon Han, Rev. Rev. B \textbf{83}, 184406 (2011).

\bibitem{PotemskiReview} M. Orlita and M. Potemski, Semicond. Sci. Technol. 
\textbf{25}, 063001 (2010).

\bibitem{GiantFaraday} Iris Crassee, Julien Levallois, Andrew L. Walter,
Markus Ostler, Aaron Bostwick, Eli Rotenberg, Thomas Seyller, Dirk van der
Marel and Alexey B. Kuzmenko, Nat. Phys. \textbf{7}, 48 (2011).

\bibitem{CalculKerr} Rahul Nandkishore and Leonid Levitov, Phys. Rev. Lett. 
\textbf{107}, 097402 (2011).

\bibitem{GorbarFaraday} E. V. Gorbar, V. P. Gusynin, A. B. Kuzmenko, and S.
G. Sharapov, Phys. Rev. B \textbf{86}, 075414 (2012).

\bibitem{Shizuya} K. Shizuya, Phys. Rev. \textbf{B} 79, 165402 (2009).

\bibitem{McCann} Edward McCann and Vladimir I. Fal'ko, Phys. Rev. Lett.%
\textbf{\ 96}, 086805 (2006).

\bibitem{RecentValues} We use the most recent values of these parameters in
the present work. See\ Jeil Jung and Allan H. MacDonald, Phys. Rev. B 
\textbf{89}, 035405 (2014).

\bibitem{ReneValidity} R. C\^{o}t\'{e} and Manuel Barrette, Phys. Rev. B 
\textbf{88}, 245445 (2013).

\bibitem{Filund} L. Wendler and V. G. Grigoryan, Physica B 245, \textbf{127}
(1998).

\bibitem{Toke} Judith S\'{a}ri and Csaba T\"{o}ke, Phys. Rev. B \textbf{87},
085432 (2013); Csaba T\"{o}ke and Vladimir I. Fal'ko, Phys. Rev. B \textbf{83%
}, 115455 (2011).
\end{thebibliography}
\end{document}